\documentclass[preprint]{aastex61}
\usepackage{natbib}
\usepackage{graphicx}
\usepackage{amsmath}
\usepackage{amsfonts}
\usepackage{amssymb} 
\usepackage[]{units}
\usepackage{nicefrac}
\usepackage{longtable}

\newcommand{\myemail}{gilhool@sas.upenn.edu}

\newcommand{\chisq}{\ensuremath{\chi^2}}

\newcommand{\teff}{\ensuremath{T_{\text{eff}}}}
\newcommand{\tirtf}{\ensuremath{T_{\text{eff, IRTF}}}}
\newcommand{\taspcap}{\ensuremath{T_{\text{eff, ASPCAP}}}}
\newcommand{\vsini}{\ensuremath{v\sin{i}}}

\newcommand{\teffe}{\ensuremath{T_{\text{eff}}}~}

\newcommand{\vsinie}{\ensuremath{v\sin{i}}~}

\newcommand{\met}{\ensuremath{[\text{M}/\text{H}]}}
\newcommand{\mete}{\ensuremath{[\text{M}/\text{H}]}~}
\newcommand{\logg}{\ensuremath{\log{g}}}
\newcommand{\logge}{\ensuremath{\log{g}}~}
\newcommand{\vrot}{\ensuremath{v_{rot}}}

\newcommand{\kms}{\ensuremath{\text{km}\,\text{s}^{-1}}}

\newcommand{\mindent}{\-\hspace{0.25cm}}

\shorttitle{M Dwarf Rotation in APOGEE}
\shortauthors{Gilhool et al.}

\begin{document}

\title{The Rotation of M Dwarfs Observed by the Apache Point Galactic Evolution Experiment}

\author{Steven H. Gilhool}
\affiliation{Department of Physics and Astronomy, University of Pennsylvania, 209 S. 33rd Street, Philadelphia, PA 19104}
\author{Cullen H. Blake}
\affiliation{Department of Physics and Astronomy, University of Pennsylvania, 209 S. 33rd Street, Philadelphia, PA 19104}
\author{Ryan C. Terrien}
\affiliation{National Institute of Standards and Technology, 325 Broadway, MC 107.02,  Boulder, CO 80305}
\affiliation{Department of Physics and Astronomy, Carleton College, One North college Street, Northfield, MN 55057}
\author{Chad Bender}
\affiliation{Department of Astronomy/Steward Observatory, University of Arizona, 933 North Cherry Avenue, Tucson, AZ 85721}
\author{Suvrath Mahadevan}
\affiliation{Department of Astronomy and Astrophysics, The Pennsylvania State University, 525 Davey Lab, University Park, PA 16802}
\author{Rohit Deshpande}
\affiliation{Department of Astronomy and Astrophysics, The Pennsylvania State University, 525 Davey Lab, University Park, PA 16802}

\correspondingauthor{Steven H. Gilhool}
\email{\myemail}

\begin{abstract}
We present the results of a spectroscopic analysis of rotational velocities in 714 M dwarf stars observed by the SDSS III Apache Point Galactic Evolution Experiment (APOGEE) survey. We use a template fitting technique to estimate \vsinie while simultaneously estimating \logg, \met, and \teff. We conservatively estimate that our detection limit is 8 km s$^{-1}$. We compare our results to M dwarf rotation studies in the literature based on both spectroscopic and photometric measurements.  Like other authors, we find an increase in the fraction of rapid rotators with decreasing stellar temperature, exemplified by a sharp increase in rotation near the M$4$ transition to fully convective stellar interiors, which is consistent with the hypothesis that fully convective stars are unable to shed angular momentum as efficiently as those with radiative cores. We compare a sample of targets observed both by APOGEE and the MEarth transiting planet survey and find no cases were the measured \vsini~and rotation period are physically inconsistent, requiring $\sin{i}>1$. We compare our spectroscopic results to the fraction of rotators inferred from photometric surveys and find that while the results are broadly consistent, the photometric surveys exhibit a smaller fraction of rotators beyond the M$4$ transition by a factor of $\sim 2$. We discuss possible reasons for this discrepancy.  Given our detection limit, our results are consistent with a bi-modal distribution in rotation that is seen in photometric surveys. 
\end{abstract}


\section{Introduction}
 The M dwarfs (main sequence stars with $M_\star \sim 0.2-0.5 M_\odot$) are the most numerous stars in the galaxy \citep{Hen1994}, but their fundamental properties are not as well measured as those of more massive stars. It has long been known that very low-mass stars tend to exhibit more rapid rotation as compared to solar-type stars, but exactly how and where along the main sequence the transition from gradual spin down to long-lived rotation occurs is not well understood \citep[e.g.][]{Houd2017}. 

 The study of rotation in M dwarfs is important as it provides a window into the structure and evolution of stellar magnetic fields and the interactions between those fields and the relatively cool atmospheres of low-mass stars. It is hypothesized that fully convective M dwarfs, generally those with spectral types M4 and later, are unable to efficiently shed angular momentum through the interaction between the stellar wind and the stellar magnetic field, and therefore spin down more slowly over time \citep{Stas2011}.  Also, some M dwarfs appear to have inflated radii, which may also be a consequence of magnetic activity (e.g. \citealt{FC2014}, \citealt{JJ2014}, \citealt{Han2017})
 
 Many current and future exoplanet surveys such as CARMENES \citep{CARMENES}, SPIRou \citep{SPIR}, IRD \citep{IRD}, MAROON-X \citep{MARX}, MEarth \citep{MEarth}, HPF \citep{Mahadevan12}, MINERVA-Red \citep{Bla15}, and NIRPS \footnote{\url{http://www.eso.org/public/teles-instr/lasilla/36/nirps/}} will target low-mass stars.  M Dwarfs are promising targets for planet searches since they are abundant, exhibit larger radial velocity and transit signals than do larger stars (all other things being equal), have close-in habitable zones, and may host many small planets (e.g. \citealt{Gai16}). On the other hand, stellar rotation and activity can degrade radial velocity precision by introducing so-called `RV jitter', so understanding the rotation properties of M dwarfs is crucial to the design of such surveys (e.g. \citealt{Des2013}). Finally, the overall rotation statistics of the M dwarf population provide important constraints on our understanding of stellar evolution at the bottom of the main sequence. 

Despite the astrophysical importance of low-mass stars, these cool and intrinsically faint stars present observational challenges.  With temperatures $\teff < \unit[4000]{K}$, these stars emit most of their radiation at infrared wavelengths, where astronomical instrumentation is less readily available and Earth's atmosphere imposes limitations on ground-based observations. The
cool temperatures of these stars enable molecules to form in their atmospheres, resulting in complex spectra with line blanketing which is difficult to model \citep{All2012}. Many of these stars are also magnetically active, resulting in spectroscopic and photometric properties that change with time. 

We present an analysis of the rotation of more than 700 M dwarfs observed as part of the The Apache Point Observatory Galactic Evolution Experiment (APOGEE; \citealt{APOGEE2017}).  Specifically, we analyze infrared spectra from the APOGEE M Dwarf Survey \citep{Des2013}, an ancillary science program  was carried out as part of the Sloan Digital Sky Survey (SDSS-III; \citealt{Eis2011}). This data set represents the largest high-resolution spectroscopic survey of M dwarfs to date, which lends unprecedented statistical power to the study of M dwarfs and their fundamental properties. In Section 2 we describe the APOGEE data set and the selection criteria used to generate the M star sample. In Section 3 we describe our template fitting approach to determining \vsini, as well as the limitations of this technique, and compare our results to those in the literature. In Section 4 we compare the distribution of projected rotation velocity to that inferred from photometric rotation periods published in the literature. Finally, in Section 5 we examine the possible implications of our results and summarize our findings. 

\section{Data Selection}
We analyzed APOGEE spectra from SDSS data release 13 \citep{DR13}, observed using the SDSS main $\unit[2.5]{m}$ telescope \citep{Gunn2006}.  The APOGEE spectrograph is a multiplexed, cryogenic, high-resolution ($R\approx 22,000$) fiber-fed instrument.  It covers the H-band ($\lambda = \unit[1.514]{\micron} - \unit[1.696]{\micron}$) across three near infrared detectors; blue ($\lambda = \unit[1.52]{\micron} - \unit[1.58]{\micron}$), green ($\lambda = \unit[1.59]{\micron} - \unit[1.64]{\micron}$), and red ($\lambda = \unit[1.65]{\micron} - \unit[1.69]{\micron}$) (\citealt{Wil2010}; \citealt{Skr2015}).  The APOGEE data pipeline produces a range of spectral products that correct for the effects of atmospheric emission and absorption and also combine spectra obtained at different epochs into single, high signal-to-noise stellar spectra in the rest frame of the star \citep{Nid2015}. In this work, we analyzed apStar spectra, which are weighted combinations of multiple spectra of each star gathered over different epochs.  For all APOGEE targets that were observed more than once, the apStar files contain two coadded spectra -- one is generated using a pixel-based weighting scheme,  and the other is generated using a global weighting scheme.  In the pixel-based weighting scheme, the $i^{\textrm{th}}$ pixel in the coadded spectrum is the signal-to-noise-per-pixel weighted combination of the $i^{\textrm{th}}$ pixels of the individual spectra.  In the global weighting scheme, the coadded spectrum is the total signal-to-noise-per-spectrum weighted combination of the individual spectra.  In most cases, there is little difference between the two.  All of the stars in our sample were observed at least twice, so we used the apStar spectra with the pixel-based weighting scheme and found no measurable difference in our results when instead using the coadded spectra with the global weighting scheme. 

We also utilized data from the APOGEE Stellar Parameters and Chemical Abundance Pipeline (ASPCAP; \citealt{Gar2016}).  Each APOGEE target has a corresponding aspcapStar file, which contains a pseudo-continuum normalized spectrum, along with parameters output from the ASPCAP pipeline.  The ASPCAP pipeline produces \teff, \met, \logg, and \vsinie measurements, in addition to estimates of up to $15$ chemical abundances for most APOGEE stars. The ASPCAP spectral libraries cover a wide range of temperatures and chemical compositions for both dwarf and giant stars.  Of interest to our work are the GK dwarf grid ($\unit[3500]{K} \leq \teff \leq \unit[6000]{K}$) and the M dwarf grid ($\unit[2500]{K} \leq \teff \leq \unit[4000]{K}$), which was added in Data Release $13$. The GK dwarf grid uses ATLAS9 models, while the M dwarf grid uses MARCS models \citep{Mes2012}.  In order to preserve continuity in the derived parameters of the GK grid, the ATLAS9 models are used in the overlapping temperature range ($\unit[3500]{K} \leq \teff \leq \unit[4000]{K}$).  This, however, produces a discontinuity in the ASPCAP parameters for the M dwarf sequence ($\unit[2500]{K} \lesssim \teff \lesssim \unit[4000]{K}$).  Applying a consistent suite of models, targeted specifically at measuring M dwarf parameters, was a primary motivation for this work.  

\begin{figure}[ht] %This is the temp comparison plot, Figure 1
\plottwo{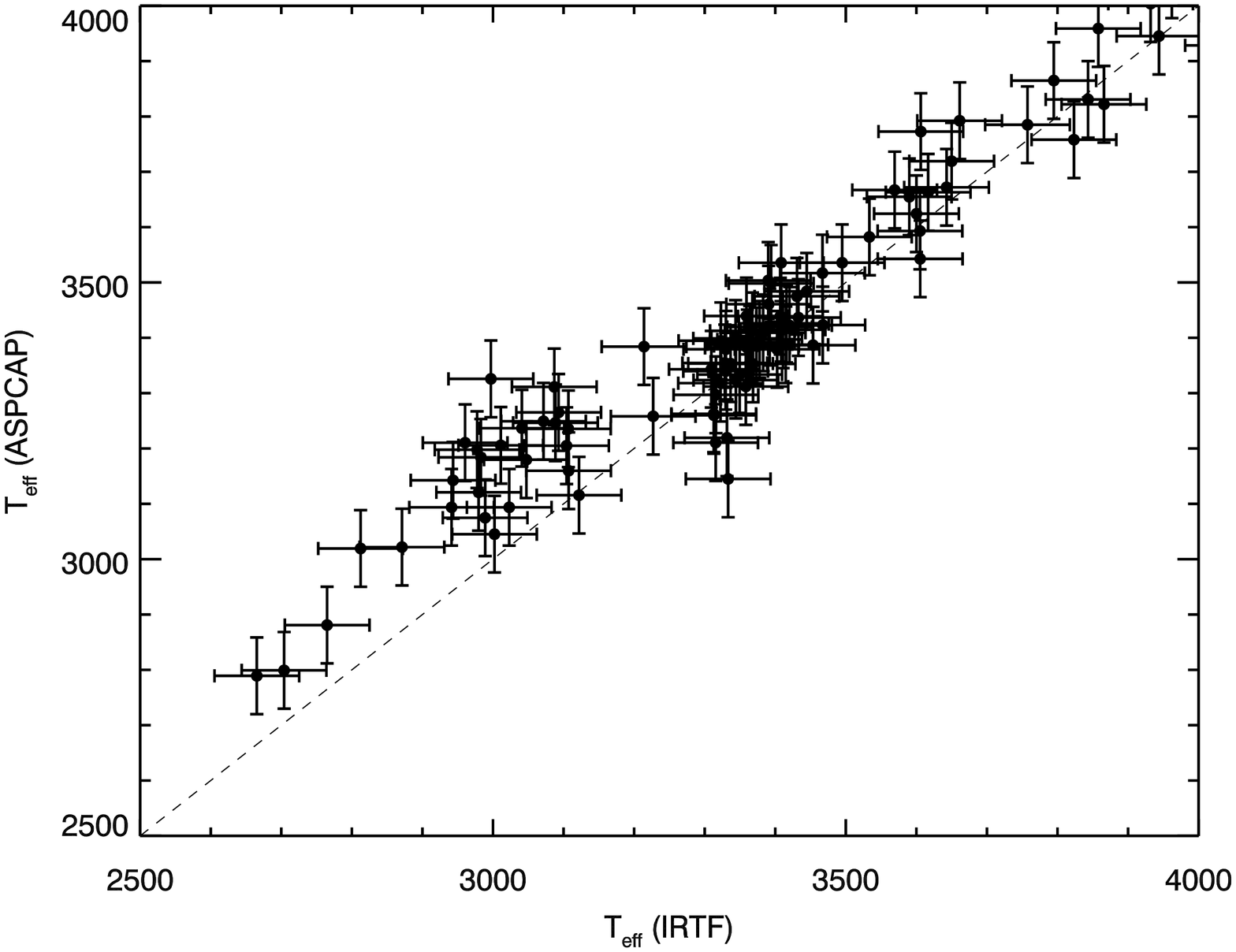}{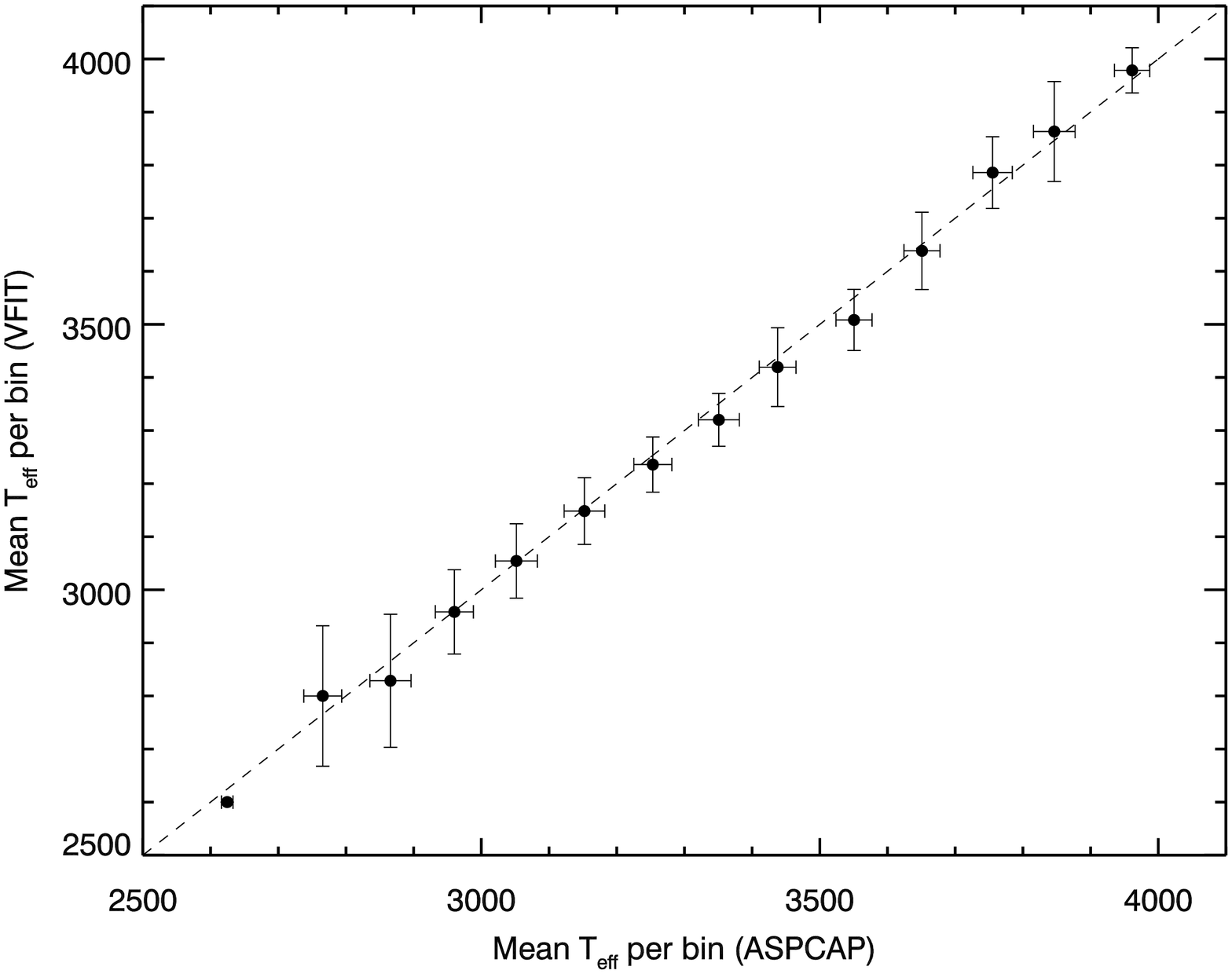}

\caption{\textbf{Left: }Comparison of ASPCAP-pipeline \teffe to values based on IRTF SpeX spectra, using $\textrm{H}_2\textrm{O}$-index relations for stars with $\teff > \unit[3300]{K}$ \citep{Mann2013}, and a V-K color-temperature relation for $\teff < \unit[3300]{K}$.  There appears to be a constant offset between the ASPCAP \teffe and the V-K color-derived \teff.  The RMS of the residuals between the ASPCAP \teffe and $\tirtf > \unit[3300]{K}$ is $\unit[70]{K}$.  The RMS of the residuals between the ASPCAP \teffe and $\tirtf < \unit[3300]{K}$ is $\unit[73]{K}$. 
\textbf{Right: }Comparison of \teffe estimates from the ASPCAP pipeline and the best-fit \teffe from template-fitting.  The points and error bars represent the mean and standard deviation of the estimated \teffe in each ASPCAP \teffe bin.  Our template-fitting procedure is not optimized to measure temperature, nor do we interpolate between the grid points, which are in increments of $\unit[100]{K}$.  Nevertheless, the data agree at the $\sim \unit[100]{K}$ level.}
\label{fig:teff_comp}
\end{figure}

In this analysis of M dwarf rotation, we made use of the ASPCAP \teffe measurements.  We tested the ASPCAP \teffe measurements against \teffe estimates based on data from the NASA-Infrared Telescope Facility (IRTF) SpeX Spectrograph \citep{Ter2015}.  These estimates are based on K-band $\textrm{H}_2\textrm{O}$-index relations from \citet{Mann2013} for $\teff > \unit[3300]{K}$, and V-K color-temperature relations for $\teff < \unit[3300]{K}$.  Generally, the temperature data are consistent, although there appears to be a constant offset between the data sets below $\teff < \unit[3300]{K}$, where the V-K color-temperature relation is used instead of the IRTF spectra.  As a secondary check, we compared our rough \teffe results from template-fitting (described in Section \ref{sec:vfit}) to the ASPCAP \teff.  Even though we use a completely different suite of theoretical stellar models, and our technique is not designed to measure \teffe precisely, the temperature data is broadly consistent at the $\sim\unit[100]{K}$ level (see Figure \ref{fig:teff_comp}).  In the absence of a \textit{de facto} standard for measuring the temperatures of cool M dwarfs, we adopt the ASPCAP \teffe values.

We chose our sample starting with the 1350 M Dwarfs observed as a part of the APOGEE M Dwarf Survey, demarcated by bit 19 in the APOGEE\_TARGET1 flag.  We then made cuts, requiring that:
\begin{enumerate}
    \item $\unit[2600]{K} \leq \teff \leq \unit[4000]{K}$ (excluded $114$ stars)
    \item SNR$ \geq 50$ (excluded $14$ stars)
    \item The star did not fail in the ASPCAP pipeline. The ASPCAP pipeline logs various warnings and flags in the APOGEE\_ASPCAPFLAGS field of the ASPCAP headers.  We cut all spectra with the `STAR\_BAD' flag, which is triggered by most failure modes (excluded $508$ stars).
\end{enumerate}

The final sample consists of $714$ stars. The number of stars, range of magnitudes, and SNR per \teffe bin are shown in Table \ref{tab:sample}. 

\section{Method}
\label{sec:vfit}

\subsection{Overview of \vsinie measurement techniques}

There are two primary methods by which \vsinie is measured.  In one method, the stellar spectrum is cross-correlated with a suite of templates (\textit{e.g.} \citealt{Del1998}, \citealt{Diaz2011}, \citealt{Rei2012}, \citealt{Hou2015}).  In this `cross-correlation' technique, \vsinie is inferred from the width of the central peak of the cross-correlation function.  Typically, observed spectra of non-rotating stars of the same spectral type are used as templates.  We tested this approach, but our sample does not include known slow rotators at the lowest temperatures.  We experimented with using theoretical templates, but our results were dominated by systematic disagreement between spectral features in the templates and those in the observations. 

The other primary method for determining \vsini, which we used in this analysis, is to make a direct, pixel-to-pixel comparison between a high signal-to-noise spectrum and a library of theoretical templates spanning a wide range of stellar parameters. The template is convolved with a rotational broadening kernel, and the kernel that produces the best fit to the data constitutes the measured value of \vsinie (\textit{e.g.} \citealt{Jen2009}). We refer to this as the `template-fitting technique' or `VFIT' technique. This technique is limited by the resolution of the data (R $\approx\unit[13]{\kms}$ for APOGEE) and also by systematic differences between spectral features found in the star and those in the spectral library. 

\subsection{Overview of template-fitting technique}
Our approach to measuring \vsinie was to forward-model the APOGEE apStar spectra using a library of theoretical templates, broadened by a theoretical broadening kernel to account for rotation.  We used BT-Settl model spectra, calculated using the PHOENIX code (\citealt{All2012}; \citealt{Bar2015}), for a grid of model parameters shown in Table \ref{tab:grid_pars}. These spectra\footnote{\url{https://phoenix.ens-lyon.fr/Grids/BT-Settl/AGSS2009/SPECTRA/}} were calculated at a wavelength grid spacing of $0.02 $\AA, and have an output grid spacing of $0.2 $\AA.  The templates with $\met \geq 0$ are not alpha-enhanced, templates with $\met = -0.5$ have $[\alpha/\textrm{H}] = +0.2$ and templates with $\met < -0.5$ have $[\alpha/\textrm{H}] = +0.4$. 

We continuum normalized the apStar and PHOENIX template spectra with our own code.  We fit a robust $5^{\textrm{th}}$-degree polynomial to the spectrum, rejecting points that deviate significantly from the estimated continuum, and then re-fit.  The process goes through $10$ iterations to arrive at our final estimate of the continuum.  It is worth noting that the continuum level of M dwarfs is often difficult to discern, due to line blanketing.  We compared our continuum-normalized apStar spectra against the continuum-normalized spectra produced by the ASPCAP pipeline and found no significant differences.

We used only the blue portion of the APOGEE spectra.  This span of approximately $\unit[620]{\AA}\,$ ($15163.52 \leq \lambda \leq \unit[15783.38]{\AA}$), was chosen because it contains stellar spectral lines that tend to agree well with the theoretical spectra, and contains ample spectral information while being small enough to keep computation times reasonable.

The fitting process proceeds iteratively.  In the first step, we fit each APOGEE spectrum to the full grid of spectral templates in order to determine the best-fit template.  The template spectra are broadened by seven different values of \vsini, ranging from $2-\unit[70]{\kms}$ in order to ensure that fast rotators are not misidentified.  Next, we divide the APOGEE spectrum into $100$-pixel segments and re-fit only the best-fit template, using a more finely spaced \vsinie grid.  Segmenting the spectrum gives us a means of excluding sections of the spectrum which are poorly modeled by the templates, and/or are contaminated by bad pixels, bright sky lines, or telluric lines, while also yielding an estimate of our single-measurement uncertainty. Finally, we used a Gaussian Mixture Model to estimate the overall best fit \vsini, properly accounting for outlier segments and our \vsinie detection limit.

\subsection{Detection Limit}
We estimated the \vsinie detection floor based on simulations with synthetic spectra, and comparisons with literature \vsini.  To simulate the detection limit, we added Gaussian noise to our templates, convolved them with a fiducial APOGEE Line Spread Function (LSF), and degraded them to the APOGEE resolution.  We artificially broadened the synthetic spectra, and then attempted to recover the \vsinie using our fitting process.  Given the perfect match between the stellar template and simulated spectra in this test, we were able to recover \vsinie down to $\vsini \sim \unit[5]{\kms}$.  However, we expect that systematic differences between the stellar templates and the actual features in the stellar spectra will degrade our ability to detect small \vsinie.

\citet{Des2013} estimated the detection floor to be $\vsini > \unit[4]{\kms}$, which is the minimum velocity at which the broadening kernel is resolved at APOGEE resolution and sampling. We adopted a more conservative detection limit here. Comparisons with a small number of available literature values of \vsinie suggested that we can recover $\vsini \sim 5$km s$^{-1}$ in most cases, but setting the detection limit at $\vsini > \unit[5]{\kms}$ would also yield a number of false detections.  Furthermore, early results using the $\vsini > \unit[5]{\kms}$ threshold showed an unexpected rising \vsinie floor with decreasing temperature.  At $\teff < \unit[3000]{K}$, nearly all \vsinie measurements were greater than $\unit[5]{\kms}$, which is statistically unlikely, given the projection effect alone. It is possible that the suite of PHOENIX models that we are using does not capture a systematic change in line width with effective temperature. For example, the model suite may not span a large enough range of surface gravity at low temperatures, resulting in artificially inflated \vsinie at low effective temperature if there is a physical correlation between \logg~and \teff. However, no such systematic effect has been found by other authors using PHOENIX templates (\citealt{Mald2017}, \citealt{Hou2015}). Tests with template grids spanning a wide range of \logge ($4.5 \leq \logg \leq 5.5$) indicate that systematic bias of \vsinie with \logge is small, and does not have a large impact on our detection threshold. 

Still, since there are no late M dwarfs in our sample that are independently known to be non-rotating, we are unable to decisively test the detection limit as a function of effective temperature.  Ultimately, we set a conservative detection limit of $\vsini > \unit[8]{\kms}$, which eliminates any inconsistencies with the literature, and admits the possibility that we cannot reliably measure $\vsini\sim\unit[5]{\kms}$ for all \teff.

\subsection{The Fitting Process in Detail}

The spectral fitting is performed using the IDL package \textit{mpfit}, which performs a least-squares optimization \citep{Mar2009}.  Each fit has three free parameters:        
\begin{enumerate}
            \item A constant line depth scale factor, $k$, to account for overall systematic differences between the observed depths of the stellar lines and those in the spectral library.  The relative intensity of the template at a given wavelength, $I(\lambda)$, is determined by the optical depth, $\tau$:
            \begin{equation*}
                I(\lambda) = e^{-\tau(\lambda)}
            \end{equation*}
            which we scale such that 
            \begin{equation*}
                I(\lambda) = e^{-k\,\tau(\lambda)}
            \end{equation*}            
            This changes the depth of all spectral lines, while maintaining their relative opacities.  Although scaling the line depths may undermine the inferred metallicity, the inclusion of this parameter significantly improved the overall quality of the fits.  We tested the code without the scaling parameter and found the comparison to literature \vsinie to be much worse.  For the three stars with higher-resolution literature \vsinie which are detectably rotating, the RMS of the residuals was $\sim\unit[0.4]{\kms}$ when we used the scaling parameter, and $\sim\unit[7]{\kms}$ without the scaling parameter.  Furthermore, our non-detections are fully consistent with the literature when using the scaling parameter, but several become inconsistent without the use of the scaling parameter.  The Pearson, Spearman and Kendall correlation coefficients all suggest that there is no correlation between the scale parameter and \vsini. 
            \item A constant multiplicative offset to the continuum level
            \item A linear term for the multiplicative offset to the continuum level
        \end{enumerate}

\begin{figure}[ht] %This is the limb dark plot, Figure 2
   \includegraphics[width=1\textwidth]{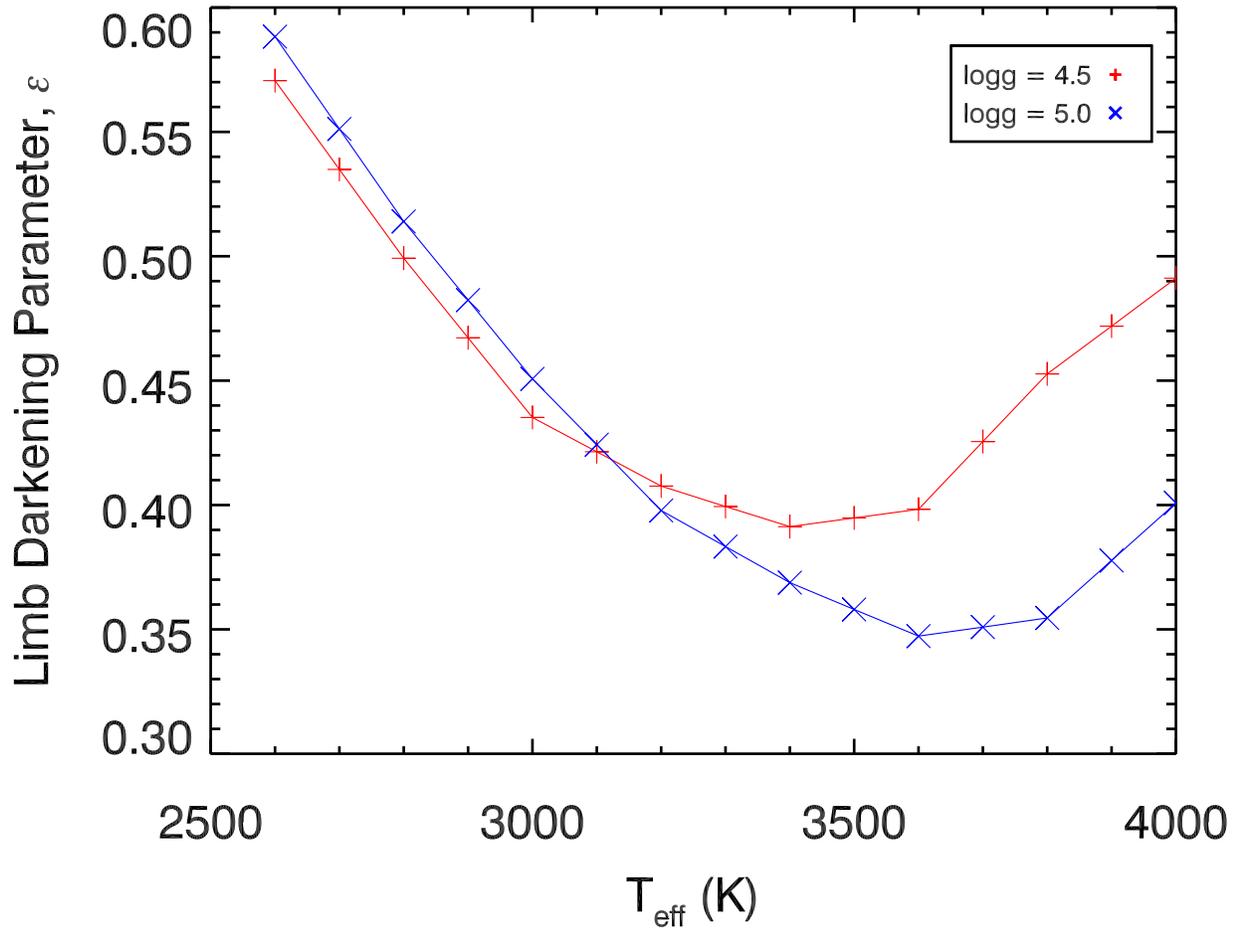}
   \caption{Limb darkening parameter values for our range of stellar parameters, calculated with \textit{jktld} code \citep{Sou2015}, using a linear limb darkening law from \citet{Cla2000}.  Values range from $\epsilon = 0.35~$ to $0.59$, with a mean value of $\epsilon = 0.44$.}
   \label{fig:epsilon}
 \end{figure}

The adjusted template is then convolved with a rotational broadening kernel whose width is determined by \vsini:
        \begin{equation*}
            G(x) = 
        \begin{cases}
        \frac{2(1-\epsilon)(1-x^2)^{1/2}+\frac{\pi\epsilon}{2}(1-x^2)}{\pi(1-\frac{\epsilon}{3})} & \lvert x \rvert < 1 \\
        0 & \lvert x \rvert > 1
            \end{cases}
        \end{equation*}
        where $x = \frac{\Delta\lambda}{\lambda_o}\cdot \frac{c}{\vsini}$ (see \citealt{Gray1992}). The limb darkening parameter, $\epsilon$, also affects the shape of the rotational broadening kernel.  We calculated $\epsilon$ using the \textit{jktld} code \citep{Sou2015} with a linear limb darkening law \citep{Cla2000}. The \textit{jktld} grids are calculated at solar metallicity and only go up to $\logg = 5.0$ so we use the same values of $\epsilon$ for the $\logg = 5.0$ and $\logg = 5.5$ templates.  Otherwise, we construct the broadening kernel using the value of $\epsilon$ corresponding to the \teff, and \logge of the current template.  These are shown in Figure \ref{fig:epsilon}. While the \citet{Cla2000} calculations give us $0.35 \lesssim \epsilon \lesssim 0.59$, other authors have used a fixed value of $\epsilon = 0.6$ (see eg. \citealt{MB2003}, \citealt{Tin1998}).  \cite{Diaz2011} caution that incorrect values of $\epsilon$ can lead to error of up to $15\%$ in \vsini.  We tested our procedure using fixed values of $\epsilon = [0.4, 0.6, 0.8]$ and found only about a $5\%$ difference in \vsinie over that range, with lower values of $\epsilon$ tending to produce higher values of \vsini.  Therefore, we adopted the \citeauthor{Cla2000} values of $\epsilon$ and conclude that our results are relatively insensitive to $\epsilon$.    
        
\begin{figure}[ht] %LSF plot, Figure 3
\includegraphics[width=1\textwidth]{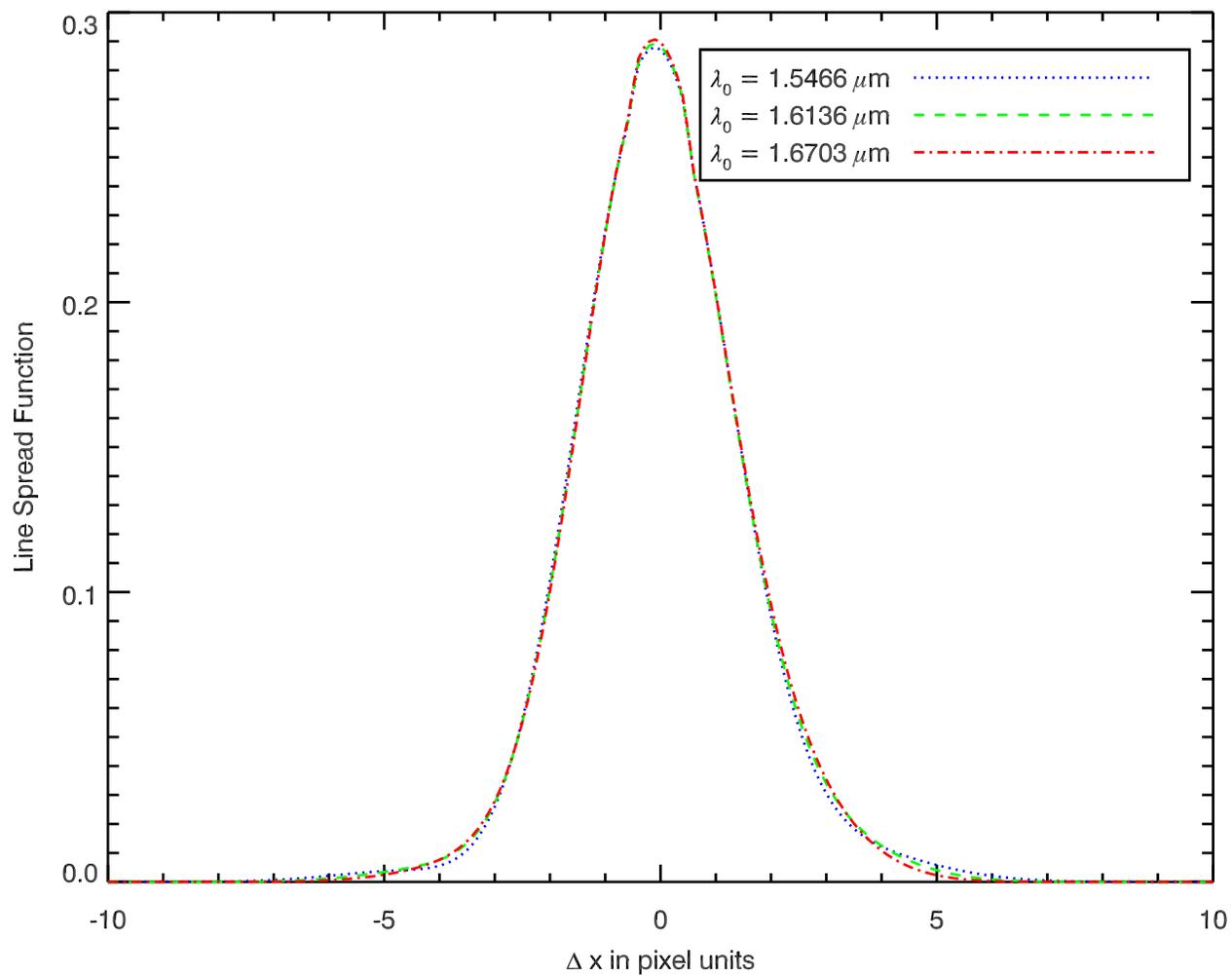}
\caption{Example of an APOGEE LSF for a single observation. The LSFs corresponding to the central wavelength of each chip (blue, green, and red) are shown. The LSF changes slightly across the detector.}
\label{fig:lsf_plot}
\end{figure}

Next, we convolve the broadened template spectrum with the APOGEE LSF in order to simulate the instrumental broadening.  We use the LSF determined by the APOGEE pipeline, which is described as a set of 26 coefficients.  The coefficients control the construction of the LSF from a series of Gauss-Hermite polynomials.  We used the IDL code, \textit{lsf\_gh}, from the SDSS idlutils package\footnote{\url{http://www.sdss.org/dr13/software/idlutils/}}.  The LSF is wavelength dependent, but only changes slightly over our spectral range (see Figure \ref{fig:lsf_plot}). We use the LSF corresponding to the central wavelength of the spectral range.

Finally, we calculate $\chisq$.  We use the APOGEE\_PIXMASK vector (HDU 3 from the apStar files) to mask bad pixels.  We consider any pixel flagged with bits $0,1,2,3,4,5,6,12, \textrm{ or } 14~$ to be a bad pixel. We do not exclude pixels based on bit $13$, which corresponds to the flag for pixels near significant telluric features.  We found no systematic biases in the analyses with and without these pixels. All flagged pixels were masked during the  $\chisq$ minimization process used to determine the best-fit \vsini.  We also mask the five pixels at either end of the spectral region to minimize edge effects resulting from convolutions in the forward modeling process.  

In the first pass of the \vsinie fitting process, the above fitting process is performed $1,890$ times (three values of \logg, $15$ values of \teff, six values of \met, and seven values of \vsini) per APOGEE star.  The result is a data cube with $\chisq$ as a function of \logg, \teff, \met, and \vsini.  At each set of \logg, \teff, and \met, we perform a cubic spline interpolation to find the minimum of $\chisq$ as a function of \vsini.  For each star, we then take the set of \logg, \teff, and \mete corresponding to the global minimum of $\chisq$ to be the best-fit template parameters.

In the second pass of the \vsini~fitting process, we fix \logg, \teff, and \mete at the previously determined best-fit values for each apStar spectrum.  We then divide each spectrum into $28$ 100-pixel chunks and repeat the fitting process, stepping through a fine grid in \vsinie ($\vsini \in [1,100]~ \kms$, in steps of $\Delta\vsini = \unit[1]{\kms}$).  The other details of the fitting process (pixel masking, limb darkening, LSF) are the same as in the first pass.  For each spectral chunk, we smoothly interpolate \chisq~ as a function of \vsinie to determine the best-fit \vsinie.  The output of this second pass is a set of $28$ \vsinie estimates per APOGEE spectrum.

We use a Gaussian Mixture Model \citep{MP2000} to robustly determine the final \vsini, based on the \vsinie estimates from the individual spectral chunks.  The model is the weighted combination of two Gaussians; one representing the `true' distribution from which the measurements are drawn, and one representing some unknown mechanism responsible for producing outliers.  The model parameters are then: $\boldsymbol{\theta} = [\,\alpha,\, \mu_{true},\, \sigma_{true},\, \mu_{out},\, \sigma_{out}\,]$, where $\alpha$ is the probability that a data point is drawn from the true distribution, and $\mu$ and $\sigma$ are the Gaussian means and widths for the true distribution and the outlier distribution, respectively.  The probability of measuring $\vsini_i$, given $\boldsymbol{\theta}$ is:

\begin{multline}
    p(\vsini_i\,|\,\boldsymbol{\theta}) = \, \frac{\alpha}{\sigma_{\textrm{true}}\sqrt{2\pi}} \ \textrm{exp}\,\bigg[-\frac{1}{2}\bigg(\frac{\vsini_i-\mu_{true}}{\sigma_{true}}\bigg)^2\bigg] \\
     +\,\frac{1-\alpha}{\sigma_{\textrm{out}}\sqrt{2\pi}} \ \textrm{exp}\,\bigg[-\frac{1}{2}\bigg(\frac{\vsini_i-\mu_{out}}{\sigma_{out}}\bigg)^2\bigg]    
\end{multline}
In principal, we could determine the model parameters by maximizing the likelihood, $L$, of the data:

\begin{equation}
  L = \prod_i^N\ p(\vsini_i|\boldsymbol{\theta})
\end{equation}

The presence of non-detections in our data, however, necessitates a
modified approach.  The numerical values of \vsini~output by our code for non-detections are
not reliable, so we cannot include them in the usual likelihood
calculation.  On the other hand, the number of non-detections places a constraint on the model, so it is necessary to incorporate them into the analysis.  We used a 
modified form of the likelihood equation developed in the statistical
field of survival analysis, which incorporates left-censored data \citep{Fei1985}.  The likelihood equation then becomes:

\begin{equation}
  L({\vsini_i}|\boldsymbol{\theta}) = \prod_i^N\ [p(\vsini_i|\boldsymbol{\theta})]^{\delta_i}\ [P(c_i|\boldsymbol{\theta)}]^{(1-\delta_i)}  
\end{equation}
In this formulation, $\delta = 1$ for detections and $\delta = 0$ for
non-detections.  $P(c_i|\boldsymbol{\theta})$ is the cumulative distribution
function, evaluated at the upper-limit for detection ($c_i = \unit[8]{\kms}$). As is common practice, we actually maximize the log likelihood, which simplifies the computation by turning the product into a summation.  We did this using the IDL code, \textit{amoeba}, which performs a downhill-simplex optimization \citep{NM1965}.  The final \vsinie measurement is the expectation value of the Gaussian defined by $\mu_{\textrm{true}}$ and $\sigma_{\textrm{true}}$. 

We used a Kolmogorov-Smirnov (KS) test to verify that the Gaussian Mixture Model was well-specified for our data.  For the vast majority of our spectra ($595/714$), the KS statistic gave at least a $90\%$ confidence that the data were drawn from the model.  Only $24$ spectra had KS statistics with less than $50\%$ confidence.  These were cases where the outlier Gaussian was narrowly peaked about a single, catastrophic outlier chunk.  As a result, the remaining $27$ chunks were all fit by the `true' Gaussian, which was not a good fit.  Upon visual inspection, however, this appears to only add a bias at the $\sim \unit[1]{\kms}$ level, which is within our errors.  

\subsection{Results}
\begin{figure}[ht] %This is the template fitting examples, Figure 4
\includegraphics[width=1\textwidth]{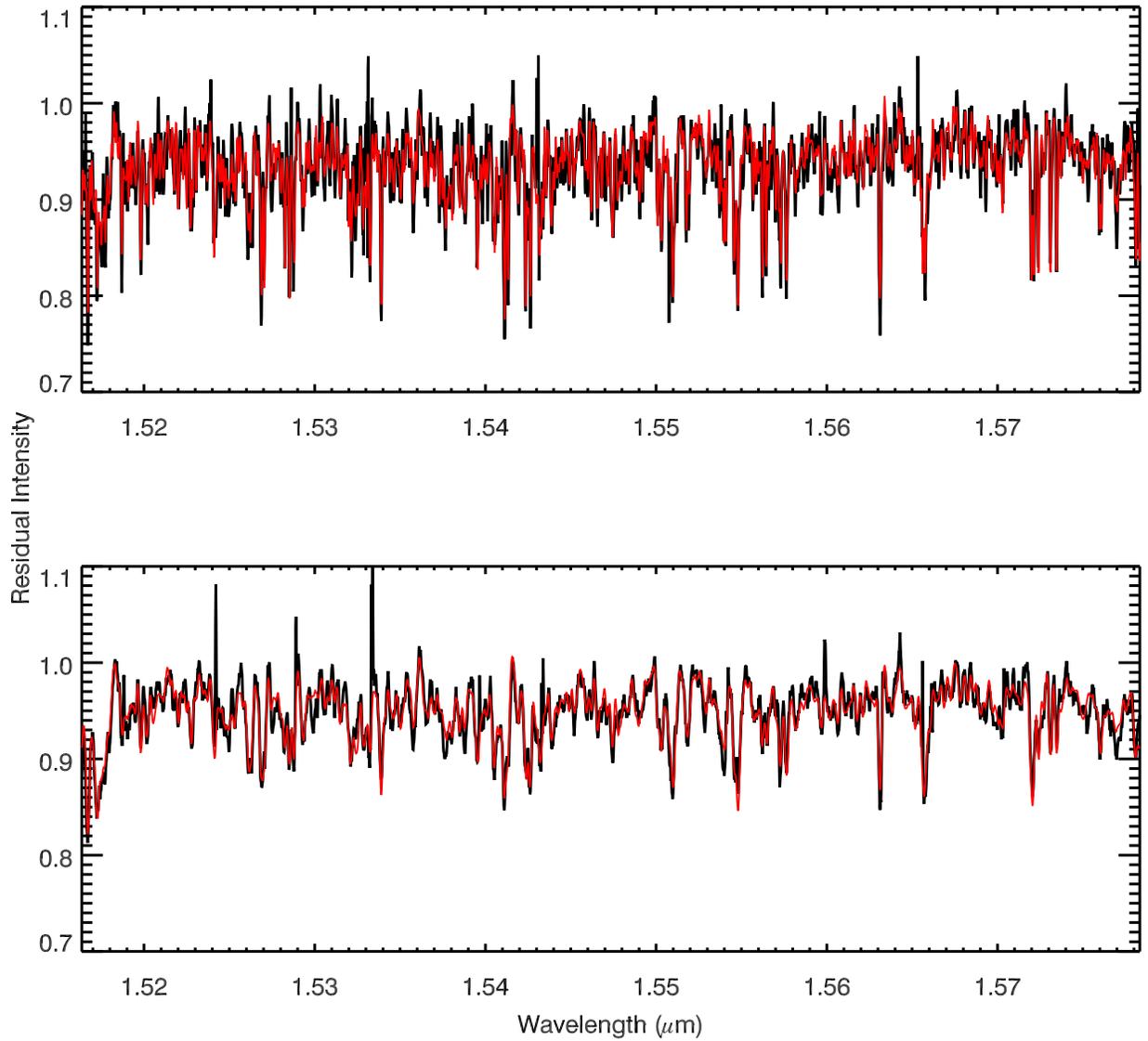}
\caption{Sample of the template fitting process.  An identical portion of the blue chip is shown for two APOGEE spectra, with the best-fit PHOENIX model in red.  These stars are similar in \teffe and \met, but the upper panel shows the spectrum of a slow rotator (non-detection), whereas the bottom panel shows a rapid rotator ($\vsini = \unit[22.6]{\kms}$).  The broadening and blending of lines due to rotation is evident.}
\label{fig:model_ex}
\end{figure}

Figure \ref{fig:model_ex} shows two example APOGEE spectra with their best-fit models superimposed in red.  The upper spectrum is considered a non-detection ($\vsini < \unit[8]{\kms}$), while the lower spectrum is that of a rapid rotator ($\vsini = \unit[22.6]{\kms}$).  The rotational line broadening in the rapid rotator is clearly discernible to the eye. These stars have similar \teff~and \met. In Figure \ref{fig:vsini_results} we examine the relationship between \vsini~ and \teff~in our sample.  Overall, our \vsinie results show a lower frequency of rapid rotators for early M dwarfs and a higher frequency of rapid rotators for late M dwarfs. This is consistent with other spectroscopic studies of M dwarf rotation in the literature (e.g. \citealt{Reiners2007}).

There are $67$ APOGEE M dwarfs with previously published \vsini, but only $16$ that we could use for comparison.  Of the $67$ published values, $9$ were from previous work on APOGEE \citep{Des2013}, $31$ were for stars with $\teff > \unit[4000]{K}$, and $11$ were for stars that failed in the ASPCAP pipeline. As shown in Figure \ref{fig:vsini_overlap} and Table \ref{table:vsini_overlap_tab}, the data are broadly consistent, aside from `2MASS J$02085359$+$4926565$,' which is from a lower resolution survey ($R\approx 19000$; \citealt{Giz2002}).  The remaining measurements all come from surveys with $R\geq 31,000$. We found that our measured \vsinie values agreed with this small number of available literature values to within approximately $\unit[3]{\kms}$ (or $0.4$km s$^{-1}$ if we only consider literature values derived from spectra with higher resolution than the APOGEE spectra).  We note that the typical reduced-\chisq of our spectral fits is much greater than $1$, due both to systematic disagreement between template and observation and  underestimated flux errors (see footnote in Table \ref{tab:sample}).  Although this could bias our results, we chose to segment the spectrum and use the Gaussian Mixture Model primarily to mitigate the effect of poor fits in certain areas of the spectrum.  Based upon the agreement with literature \vsinie values, and simulations we have carried out using simulated APOGEE spectra, we are confident that our \vsinie estimates are robust at the level of the total uncertainties described below.

We estimated our \vsinie uncertainty for each star as the quadrature sum of three sources of error. Based on the results of the Gaussian Mixture Model fits to the chunk-based \vsinie estimates, we scale $\sigma_{true}$ by the square root of the number of non-outlier chunks ($\sqrt{\textrm{floor}(\alpha \cdot 28)}$) as an estimate of the single measurement uncertainty in our fitting process. We include the $0.4$~km s$^{-1}$ uncertainty in the absolute scale of \vsini~based on the comparison to literature values, and also a 1 km s$^{-1}$ uncertainty resulting from systematic template mismatch due to the coarse sampling in \mete and \teffe in our template grid.  This was based on tests where fits for \vsini~were forced with an incorrect template, off in \teff, \met, and/or \logge by up to two grid points. As shown in Figure \ref{fig:mismatch}, the impact on \vsini~is small, approximately 1 km s$^{-1}$. 

Additionally, we performed a similar test to ensure that our two-step fitting process was reliable.  In order to estimate the impact of choosing the wrong best-fit template in the first iteration, we created a broadened synthetic spectrum from a PHOENIX model and ran the fit using templates with parameters which were off by $\pm 1$ grid point in \mete and \logg, and off by up to $\pm \unit[300]{K}$ in \teff.  The resulting \vsinie error was similarly about $1\kms$. We also investigated the relationship between the signal-to-noise of our spectra and the estimated \vsini. Since cooler stars will tend to be fainter, and therefore have lower signal-to-noise spectra, a bias in our analysis could artificially inflate the trend we see of increasing rotation with decreasing \teff. We simulated this effect and found no such bias. As shown in Figure \ref{fig:sn}, on average we recover the known \vsini~in simulated spectra with a fidelity of better than 1 km s$^{-1}$ even for the faintest stars in our sample.  We estimate a total \vsinie measurement uncertainty between 1 and 3 km s$^{-1}$ for the majority of our targets. The full results of the fitting process are reported in Table \ref{table:myvsini}.

\section{Comparison with Rotation Periods from Photometry}

In this analysis, we used spectroscopic data to measure the projected rotational velocity, but it is also possible to infer rotation periods (without the projection effect) from periodic photometric variations due to star spots rotating across the stellar surface (\textit{eg.} \citealt{Irw2011}, \citealt{Mc2014}, \citealt{New2016}).  If the stellar radius is known, the rotation period, $P$, can be converted to an equatorial rotational velocity: $\vrot = \frac{2 \pi \textrm{R}_\star}{P}$.  \citet{New2016}, for example, uses the mass-radius relation from \citet{Boy2012} to estimate the stellar radius and convert period measurements from the MEarth survey to rotational velocities. Spectroscopic and photometric methods are sensitive to different regimes of rotation velocity. While our \vsini~detection limit means that we are typically not sensitive to rotation periods longer than $\sim 1-2$~days, photometric surveys are sensitive to periods shorter than tens of days. However, comparisons can be made between the two techniques. 

We sought to make a quantitative comparison between our spectroscopic projected rotation velocities and the rotation velocities derived from photometric periods in the literature.  In Figure \ref{fig:per_vsini_comp}, we directly compared our \vsinie to the $v_{rot}$ from \citet{New2016}.  Since \vsinie is the minimum possible rotational velocity, the shaded region ($\vsini > v_{rot}$) is non-physical.  Even though we are comparing \vsinie to $v_{rot}$, we do expect most data points to fall near the one-to-one correspondence line, because the distribution of $\sin{i}$ peaks sharply at $\sin{i} = 1$ (assuming the spin axes are randomly oriented). Specifically, half of the stars should have $\sin{i} \geq 0.86$, and that is in fact the case for the $10$ detections in the sample. A KS test yields a $77\%$ confidence that the data are consistent with randomly distributed spin axes.
Therefore, the direct comparison of \vsinie and $v_{rot}$ for stars with both measurements appears consistent, given the uncertainties. 

We further sought to compare the distributions of rotational velocity as a function of effective temperature.  In this case, photometric studies suggest a larger proportion of slowly rotating late-type stars that may not be present in our data.  \citet{New2016} compared their rotation period measurements from the MEarth survey to \vsinie measurements from \citet{Del1998}, \citet{MB2003} and \citet{Brow2010}.  Both \citeauthor{Del1998} and \citeauthor{MB2003} found that approximately $50\%$ of mid M dwarfs were detectably rotating, while \citeauthor{Brow2010} reported a rotation fraction of $30\%$.  Assuming a stellar radius of $0.2 R_{\odot}$, the \vsinie detection limit of those surveys ($\sim \unit[3]{\kms}$), corresponds to rotation periods of $P\lesssim \unit[3.3]{days}$.  \citet{New2016}, however, finds just $18\pm2\%$ of mid M dwarfs in their sample would be detected as rotating in the aforementioned \vsinie studies.   

We calculated the fraction of rotators ($\vsini > \unit[8]{\kms}$) as a function of effective temperature for a number of data sets. In Figure \ref{fig:fast_rot_frac}, we show the fraction of rotators in temperature bins.  The error bars bracket $90 \%$ confidence intervals assuming binomial statistics \citep{Geh1986}.  The red points are from earlier \vsinie analyses (see \citealt{Rei2012} and references therein). The literature \vsinie are binned by spectral type, and mapped onto the temperature scale based upon relations between stellar temperature and spectral type, derived from Table 5 of \citet{Pec2013}.

The shaded regions in Figure \ref{fig:fast_rot_frac} are estimated rotation fractions based upon two sets of photometric period data.  We simulated the expected \vsinie distributions and the resulting rotation fractions using a Monte-Carlo analysis. The green region is based on rotation periods from Kepler photometry \citep{Mc2014}, and the blue region is based on rotation periods from MEarth photometry \citep{New2016}.  The procedure was performed as follows:  
\begin{enumerate}
    \item \textbf{We bin the data in bins of $\unit[100]{K}$ for $2600 \leq \teff \leq \unit[4000]{K}$.}\\  
    \mindent \textit{Kepler:} We used the \teffe data from Table $3$ of \citet{Mc2014}.\\
    \mindent \textit{MEarth:} No \teffe estimates were provided, so we used the stellar mass estimates from \citet{New2016} and inferred a temperature from the stellar mass using a relation based on \citet{Pec2013}.

    \item \textbf{We generate a realistic set of rotational velocities, $v_{rot}$, for each temperature bin.}\\  
    \mindent \textit{Kepler:} We used the quoted rotation periods, $P$, and errors, $\sigma_{P}$.  We had to infer a radius for each star from its temperature.  We estimated the stellar radii using an empirical relation from Table $1$ of \citet{Mann2015}.  This relation is quoted as having an uncertainty of $13.4\%$, which we applied to the radius estimates.\\
    \mindent \textit{MEarth:} We used the $v_{rot}~$ from \citet{New2016}, which were calculated from their period and radius measurements.  We include a $13\%$ uncertainty in \vrot, based on an assumed $10\%$ uncertainty in the other two quantities.  The measurements are tagged with a quality flag: (A) and (B) for confident detections, (U) for possible detections, and (N) for non-detections.  We use only the stars with (A) and (B) flags.
    \item \textbf{We generate a corresponding set of $\sin{i}$ values.}\\
        Assuming a random distribution of spin axis orientations, the probability density function of $\sin{i}\in[0,1]$ is:
        \begin{equation*}
            p(\sin{i}) \propto \frac{\sin{i}}{\sqrt{1 - \sin^2{i}}}
        \end{equation*}
        This is the same as a uniform distribution of $\cos{i}$.  We generated random inclinations by generating a uniform distribution of $\cos{i}$ and transforming using $\sin{i} = \sqrt{1-\cos^2{i}}$.
    \item \textbf{We combine (2) and (3) to simulate $\vsini = v_{rot}\cdot\sin{i}$, and calculate the rotation fraction.}\\
    Since our simulated populations of \vsinie were generated from the stars with detected periods, we had to account for the number of stars observed which had no detectable period. 
    
    \mindent \textit{Kepler:}  \citet{Mc2014} noted that the periodic fraction of stars with $\teff < \unit[4000]{K}$ is $83\%$, so we scaled the total number of stars in each bin accordingly.  They report completeness of $\sim 95\%$, which we take to be $100\%$ for our purposes.\\
    \mindent \textit{MEarth:} We added the stars with (U) and (N) flags in each bin as non-detections.  We also scaled up the number of detections to account for the difference in sensitivity between Kepler and MEarth.  Figure \ref{fig:per_amplitudes} shows the cumulative distribution of period detections as a function of amplitude for Kepler (in black) and MEarth (in red).  Approximately $40\%$ of the Kepler detections fall below the minimum sensitivity of MEarth, so we scaled up the number of detections to account for the expected missing detections.
    
\end{enumerate}

In Figure \ref{fig:fast_rot_frac} we show the results of a Monte Carlo simulation to determine the expected rotation fraction as a function \teffe based on the Kepler and MEarth photometric periods. 
Although the spectroscopic and photometric results appear consistent for the sample of stars with both measurements (see Figure \ref{fig:per_vsini_comp}), we found that the distribution of rotations as a function of temperature exhibited some disagreement for stars later than $\sim \textrm{M}3$. The spectroscopically-derived \vsinie estimates seem to suggest a higher fraction of late, fully-convective M dwarfs are rapidly rotating ($\vsini > 8$km s$^{-1}$) as compared to the photometrically-derived rotation periods.  It is interesting to note that \vsinie studies, both ours and those in the literature, tend to show a sharp change in behavior at the $\textrm{M}4$ transition, while the photometric studies show rotation fractions increase much more gradually along the M dwarf sequence.

\section{Discussion and Conclusions}

We present results from an analysis of high-resolution spectroscopic observations of 714 M dwarfs obtained by the SDSS APOGEE survey. We derive estimates of projected rotation velocity, \vsini, by fitting a large suite of rotationally broadened theoretical templates to these observations. We analyze these spectra in chunks and use a Gaussian Mixture Model approach to estimate our measurement uncertainty based on individual fits to 28 chunks of each spectrum. For each of our targets, we estimate an overall uncertainty of 1 to 3 km s$^{-1}$ and find good agreement between our measurements and a small number of previously published values. 

Through a Monte Carlo simulation, we attempt to make a direct comparison between the overall distribution of our projected rotational velocities and the photometric periods published in the literature. While we do find broad agreement in that rotation fraction increases with lower stellar temperature in all data sets, for $\teff<3200$~K we see a rotation fraction that is a factor of $\sim2$ higher than the rotation fraction that would be inferred from the MEarth photometry (see Figure \ref{fig:fast_rot_frac}). There are a number of factors which could explain this tension.  One important factor concerning the distribution of rotational velocities is the stellar age of the population.  We have not made any explicit selection cuts based on age here, though it is possible that the reduced proper motion criteria used in \citet{Des2013} to originally select our sample could induce an age bias.  The APOGEE M Dwarf sample was assembled using two different catalogs with two different proper motion cuts.  Part of the APOGEE sample was selected with $\mu > 150$~mas/yr, and part was selected with $\mu > 40$~mas/yr.  Similarly, \citet{New2016} noted that MEarth proper motion cuts of $\mu > 150$~mas/yr likely excluded some kinematically cold stars, which would tend to be younger and more rapidly rotating.  We applied the kinematic age estimation method presented in \citet{New2016} to our APOGEE targets and found a slightly larger proportion of likely young stars than in the MEarth sample.

There are at least two factors which could contribute to the high rotation fractions for late M dwarfs seen in our \vsinie analyses.  The first is the possibility that the detection threshold is higher than assumed.  If, for example, our detection threshold were $\vsini = \unit[12]{\kms}$, then the fraction of rotators at $\teff<3200K$ would shift down in Figure \ref{fig:fast_rot_frac} and would be almost entirely consistent with the inferred rotational velocities from MEarth photometry. Another systematic error which may bias our results is binary contamination.  We have no particular mechanism for filtering out binaries. 

One possible systematic effect arising from the photometric analyses is aliasing.  It is possible that some of the reported periods are harmonics of the true rotation period.  In \citet{New2016}, for example, the authors addressed this by listing alternate rotation periods for stars which disagreed with the \vsinie from the literature.  Also, being a ground-based survey, MEarth is sensitive to aliasing at $1$-day periods.  Similarly, the photometric periods may not be strictly indicative of rotation periods, rather a combination of rotation and any periodic behavior intrinsic to spots themselves. This effect is difficult for us to quantify as it is unknown what the characteristic spot evolution timescales are for cool stars. Finally, we may be underestimating the rotation fraction by counting all stars without periods as non-detections.  It is very likely, for example, that some of the MEarth stars with the (U) flag are rotating.  In fact, the authors note that even the (N) flag does not mean that the star is not rotating; it simply means that they were unable to detect a periodic modulation. We also note that even though our sample represents the largest set of high-resolution spectra of M dwarfs, there are fewer than $10$ stars in each of the three coolest temperature bins.

One interesting feature of our analysis is the comparatively small fraction of non-detections in our sample below $\teff \sim \unit[3000]{K}$. Approximately $65\%$ of the these coolest dwarfs in our sample are non-detections of rotation given our conservative 8 km s$^{-1}$ detection limit. This is consistent with the trend toward faster rotation at later spectral types observed in earlier \vsinie analyses.  It has been suggested based on photometric data that the rotation periods of low-mass stars may be bi-modal (e.g. \citealt{Irw2011},\citealt{Mc2014}, \citealt{New2016}). For example, Figure 15 of \citet{New2016} shows a substantial population of stars with masses less than 0.3M$_\Sun$ and rotation periods longer than 10 days. These would all be non-detections in our analysis, even if we were to assume a less conservative \vsinie detection limit.

\begin{figure}[ht] %vsini results, Figure 5
   \includegraphics[width=1\textwidth]{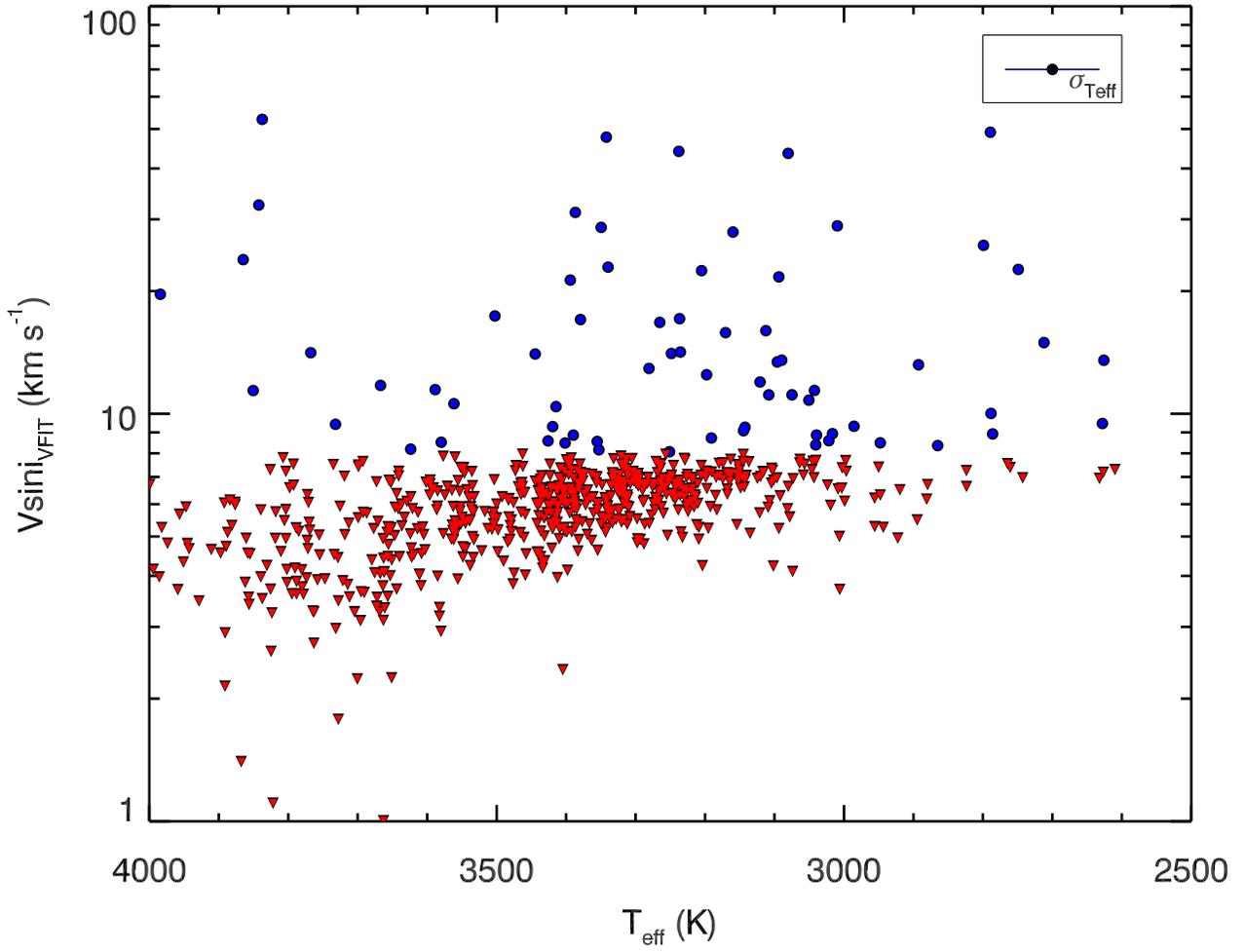}
   \caption{\vsinie results from our template fitting approach.  The red downward facing triangles represent non-detections ($\vsini < \unit[8]{\kms}$) and are plotted at the calculated value for visualization purposes.  The blue circles are measured values of \vsini.  Effective temperatures are those from Data Release 13 of the ASPCAP pipeline.  The typical error for \teffe is shown in the legend.  Uncertainties in \vsinie range from $1-\unit[3]{\kms}$.}  
   \label{fig:vsini_results}
 \end{figure}

\begin{figure}[ht] %Comparison with literature vsini, Figure 6
\includegraphics[width=1\textwidth]{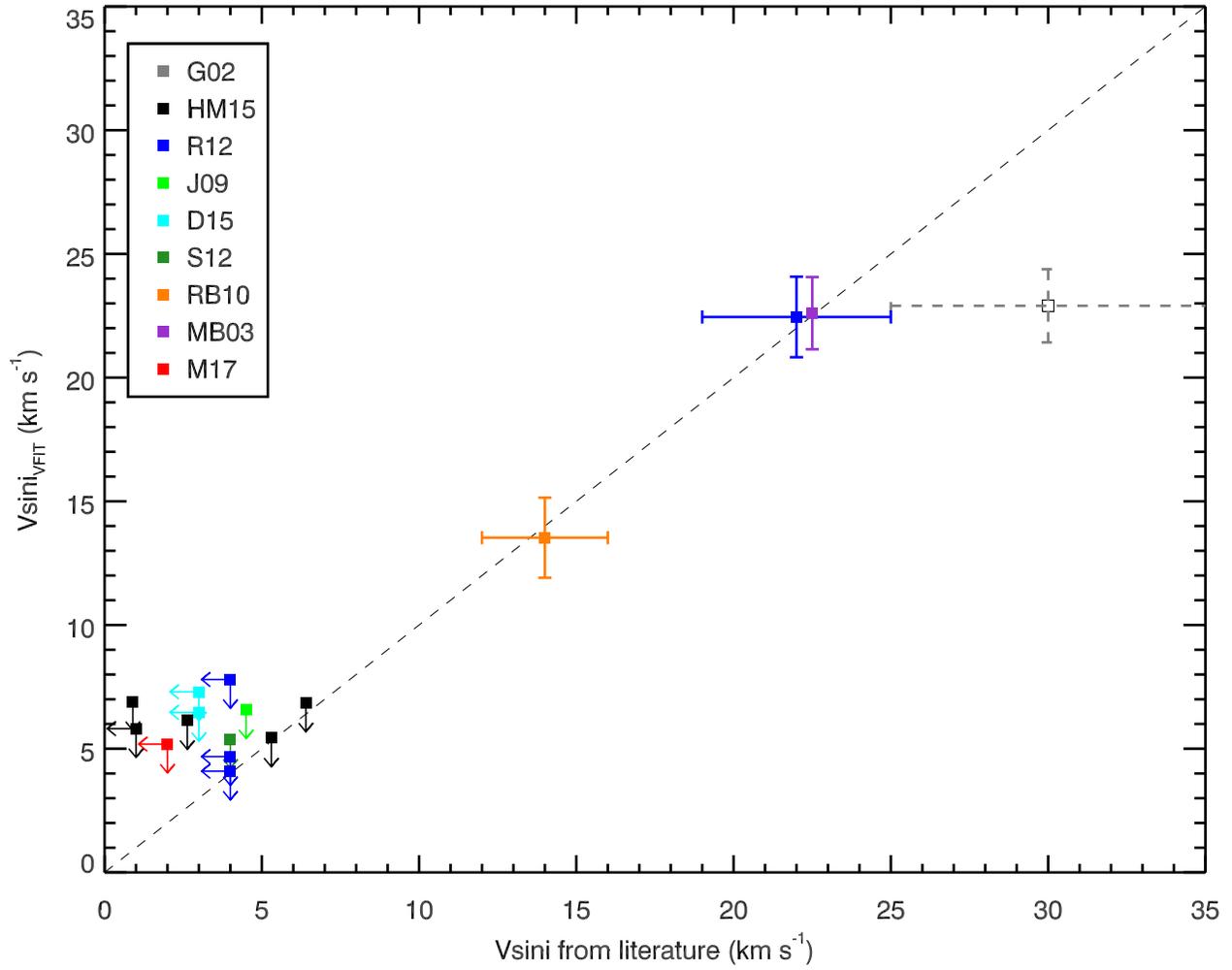}
\caption{Comparison of our \vsinie measurement with literature values, color-coded by reference.  See Table \ref{table:vsini_overlap_tab} for full citations.  The arrows denote upper limits for non-detections.  The fastest literature rotator, `2MASS J$02085359$+$4926565$,' (unfilled square with dashed error bars) is from a lower resolution survey ($R\approx 19000$; \citealt{Giz2002}).  Our results are entirely consistent with the literature.}
\label{fig:vsini_overlap}
\end{figure}

\begin{figure} %Template mismatch test, Figure 7
\plotone{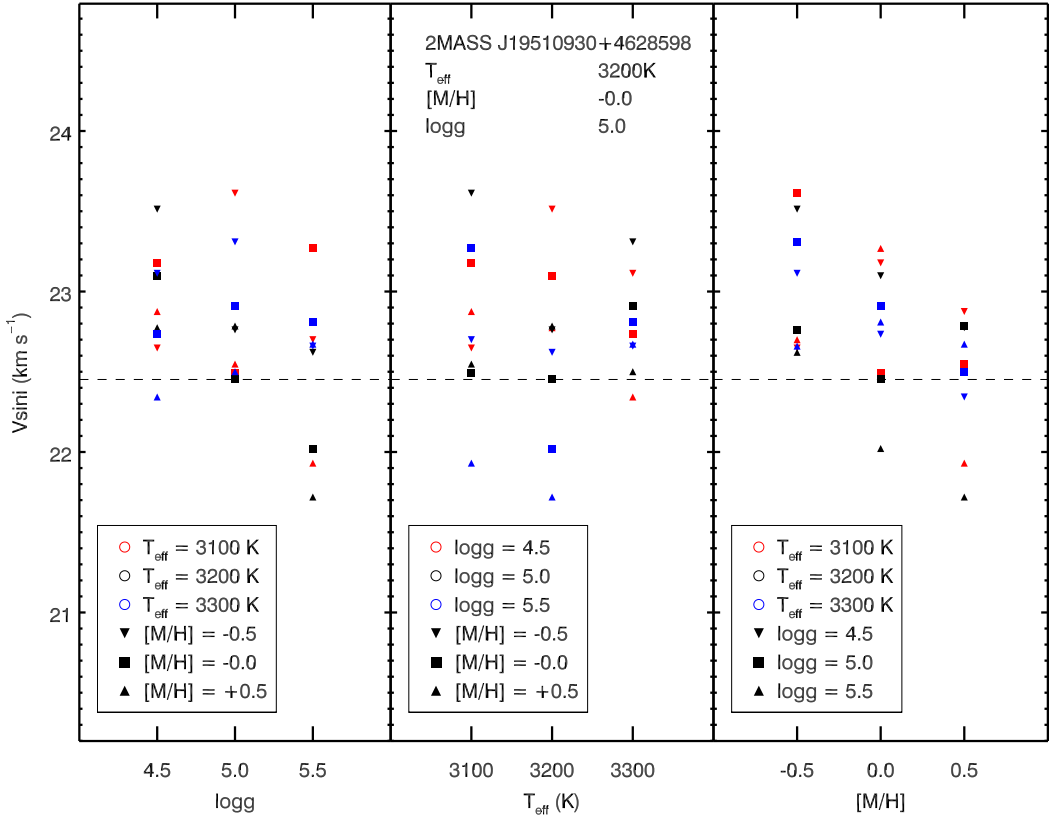}
\caption{Test of the effect of misidentifying the best-fit PHOENIX template.  We fit the set of stars with previously published \vsinie using a mini-grid of templates adjacent to the best-fit template.  The results for `2MASS J19510930+4628598' are shown here.  The dashed line shows our reported value of \vsinie and each point is a \vsinie measurement, corresponding to a point in the mini-grid.  In each panel, \vsinie is plotted against one grid parameter, while the other two grid parameters are encoded in the shape and color of the plot symbol (denoted in the legend).  The RMS of the residuals is $0.53 \kms$, and the maximum deviation is $\Delta \vsini = \unit[1.2]{\kms}$.  We assume a $1\, \kms$ systematic error from possible template mismatches.}
\label{fig:mismatch}
\end{figure}

\begin{figure} %Flux error test, Figure 8
\label{fig:fluxerror}
\plotone{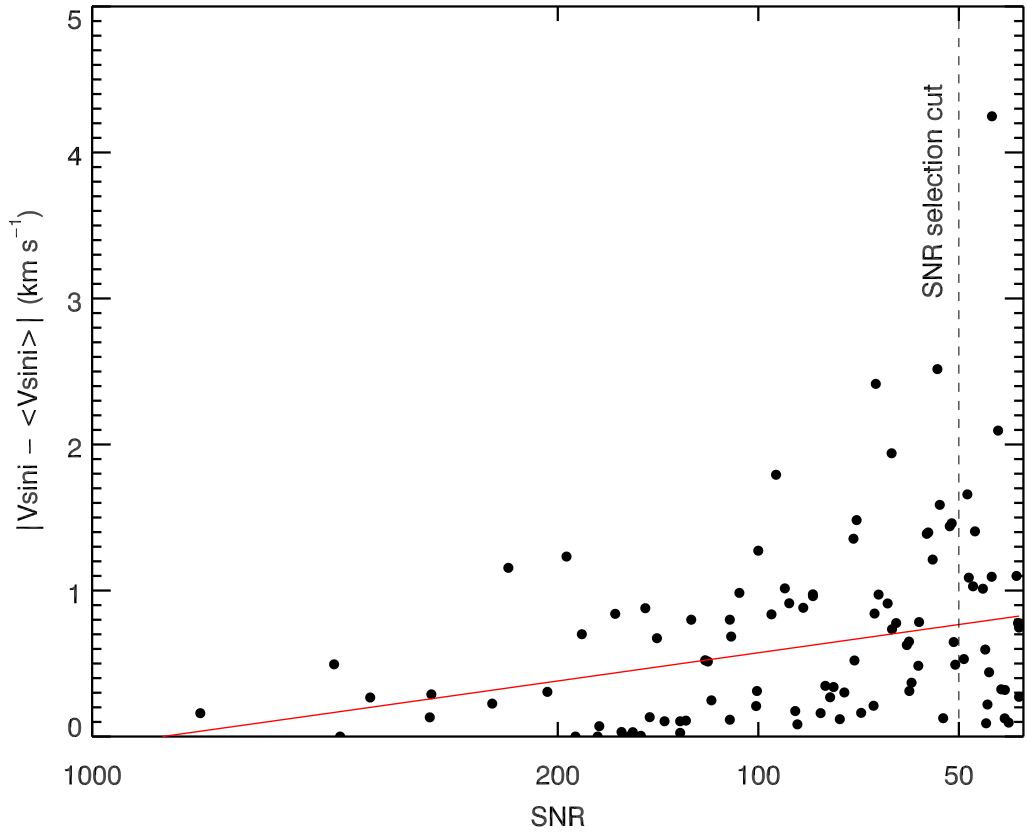}
\caption{Deviation in \vsinie as a function of SNR.  We added Gaussian noise to the spectra of several rotators and measured the \vsinie of the degraded spectra.  A robust linear fit yields an average deviation of about $0.75 \,\kms$ at a SNR of 50, which is the minimum allowed in our sample.  We also performed this procedure on non-detections.  A small minority of trials ($2\%$ at $50 < \textrm{SNR} < 70$) yielded \vsinie above the detection limit.  There are only $37$ stars in our sample with SNR in this range, so it is unlikely that any are false positives due to flux error alone.}
\label{fig:sn}
\end{figure}

\begin{figure}[ht] %Direct comparison of vsini to photometric rotational velocity, Figure 9
\includegraphics[width=1\textwidth]{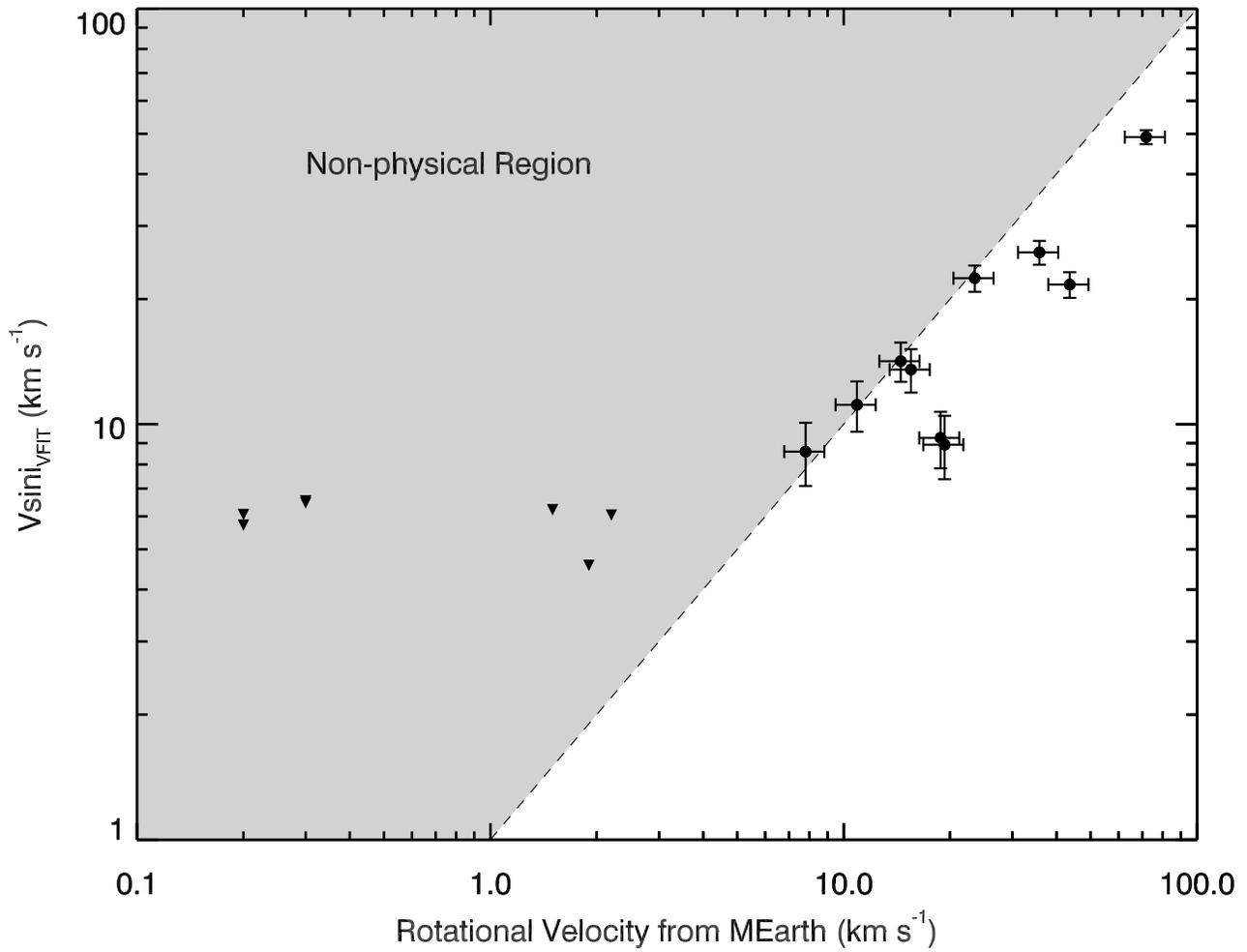}
\caption{Comparison of our \vsinie measurement versus equatorial velocities calculated from photometric periods in \citet{New2016}.  Since \vsinie is the minimum possible equatorial velocity, the gray shaded region is non-physical. The downward triangles are non-detections in our analysis, and are therefore not inconsistent with the MEarth data.}
\label{fig:per_vsini_comp}
\end{figure}

\begin{figure}[ht] %Rotation fraction vs. Teff plot, Figure 10
\includegraphics[width=1\textwidth]{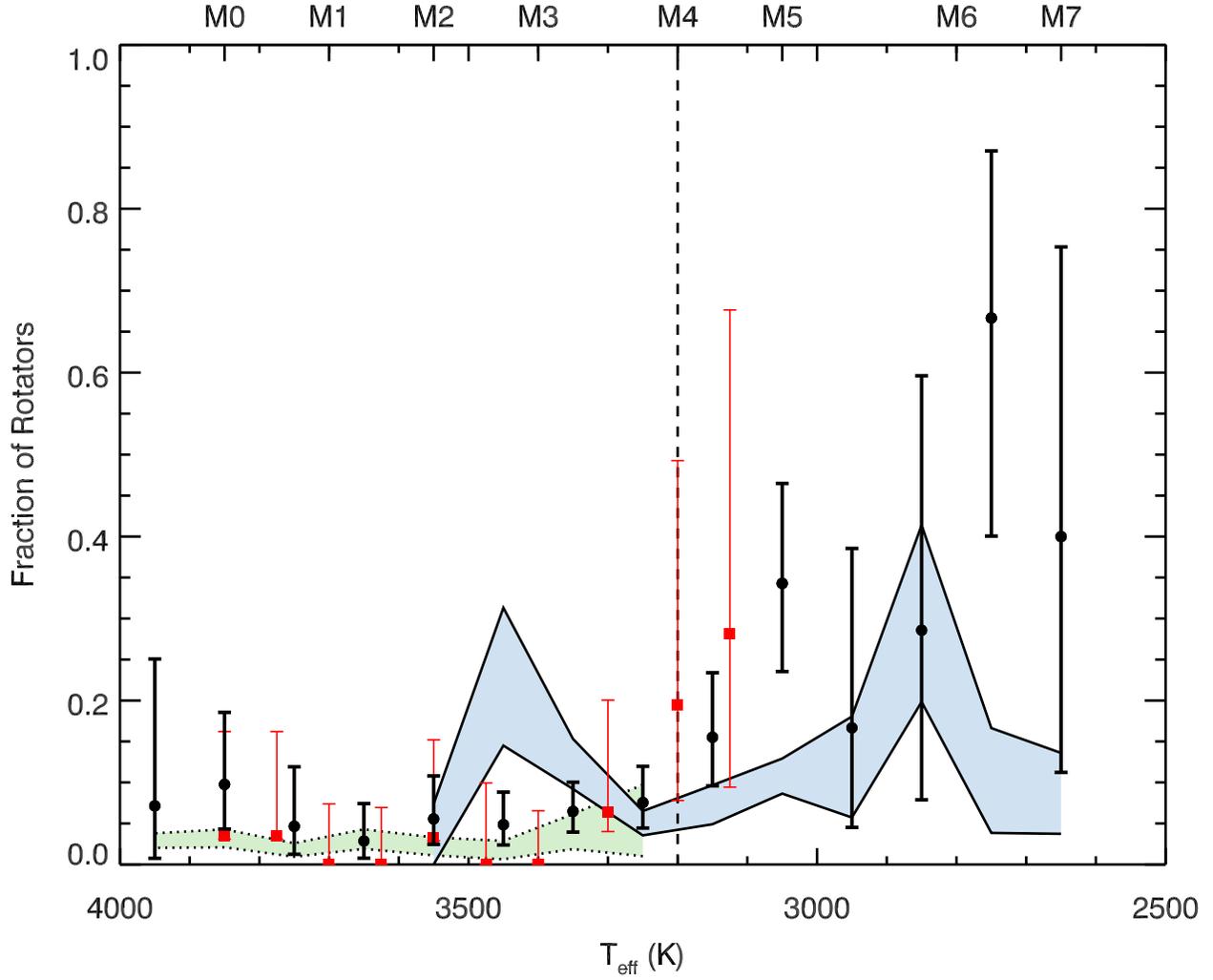}
\caption{Fraction of rapid rotators as a function of effective temperature.  The black points (circles) are based on our $\vsini$ measurements, with $90 \%$ confidence intervals from binomial statistics.  The red points (squares) are based on \vsinie from literature sources (\citealt{Rei2012} and references therein) also with $90 \%$ confidence intervals from binomial statistics.  These are binned by spectral type, and mapped onto our \teffe scale.  The green region with the dotted outline is the $90 \%$ confidence interval of simulated rotation fractions based on rotation periods measured from Kepler photometry \citep{Mc2014}.  The blue region with the solid outline is the $90 \%$ confidence interval of simulated rotation fractions based on rotation periods measured from MEarth photometry \citep{New2016}.}
\label{fig:fast_rot_frac}
\end{figure}

\begin{figure}[ht] %Photometric variability amplitude sensitivity, Figure 11
\includegraphics[width=1\textwidth]{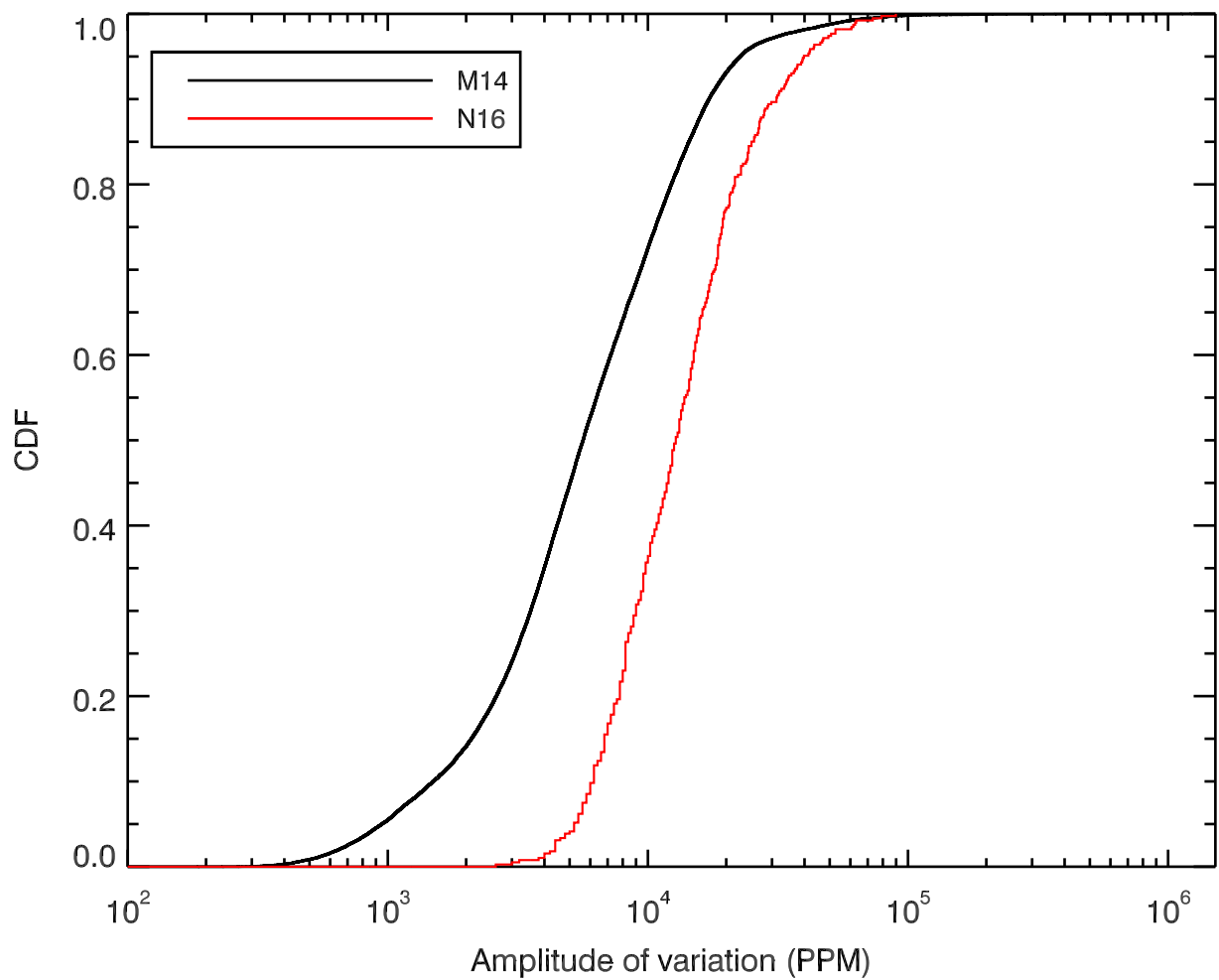}
\caption{Comparison of Kepler and MEarth sensitivity to periodic photometric variations.  Approximately $40\%$ of the periods detected in \citet{Mc2014} have amplitudes that are too small to be detected and classified as (A) or (B) rotators in \citet{New2016}.}
\label{fig:per_amplitudes}
\end{figure}

Overall, our \vsinie results show a low frequency of rapid rotators for early M dwarfs and a high frequency of rapid rotators for late M dwarfs, with a sharp transition which roughly coincides with the $\textrm{M}4$ transition to fully-convective stellar interiors (see Figure \ref{fig:vsini_results}). This is consistent with other spectroscopic studies of M dwarf rotation in the literature (e.g. \citealt{Reiners2007}). As shown in Figure \ref{fig:per_vsini_comp}, we find good agreement for individual targets between our \vsinie estimates and published photometric periods. Since $\sin{i}$ can not be larger than 1, a given equatorial rotation velocity derived from a photometric period and stellar radius sets a physical limit on \vsini. Within a sample of 19 stars with photometric periods for which we measured \vsini, we found no examples non-physical rotational velocities.  Through a Monte Carlo simulation, we make a direct comparison between the overall distribution of our projected rotational velocities and the photometric periods published in the literature. While we do find broad agreement in that rotation fraction increases with lower stellar temperature in all data sets, for $\teff<3200$~K we see a rotation fraction that is a factor of $\sim2$ higher than the rotation distribution that would be inferred from the MEarth photometry (see Figure \ref{fig:fast_rot_frac}). This rotation fraction depends both on the detection limit in our analysis, which we conservatively set at 8 km s$^{-1}$ as well as the fraction of stars with a given rotation rate that exhibit detectable photometric variability. While we attempt to quantify this effect by comparing the distribution of photometric amplitudes found in the MEarth survey to those in the Kepler survey (which has substantially better photometric precision), it is possible that a significant population of rotating stars without photometric variability remains. While the fraction of non-detections at $\teff<3000K$ depends sensitively on our assumed \vsini~detection limit, the fraction of rotators in our sample appears consistent with the bi-modal photometric period distribution seen in the MEarth and Kepler studies.

\acknowledgements

We would like to thank John Bochanski, Matthew Shetrone, Keivan Stassun, and Nick Troupe for their insights into the details of the APOGEE pipeline, and suggestions regarding possible biases in this work. We would also like to thank an anonymous referee for his or her comments that helped to improve this manuscript.  

This work was supported in part by the Ella N. Pawling Endowment. This research has made use of NASA's Astrophysics Data System and the SIMBAD database,
operated at CDS, Strasbourg, France \citep{Wen2000}. Funding for SDSS-III has been provided by the Alfred P. Sloan Foundation, the Participating Institutions, the National Science Foundation, and the U.S. Department of Energy Office of Science. The SDSS-III web site is http://www.sdss3.org/. SDSS-III is managed by the Astrophysical Research Consortium for the Participating Institutions of the SDSS-III Collaboration including the University of Arizona, the Brazilian Participation Group, Brookhaven National Laboratory, Carnegie Mellon University, University of Florida, the French Participation Group, the German Participation Group, Harvard University, the Instituto de Astrof\'isica de Canarias, the Michigan State / Notre Dame / JINA Participation Group, Johns Hopkins University, Lawrence Berkeley National Laboratory, Max Planck Institute for Astrophysics, Max Planck Institute for Extraterrestrial Physics, New Mexico State University, New York University, Ohio State University, Pennsylvania State University, University of Portsmouth, Princeton University, the Spanish Participation Group, University of Tokyo, University of Utah, Vanderbilt University, University of Virginia, University of Washington, and Yale University.

\software{idlutils \url{http://www.sdss.org/dr13/software/idlutils/}, mpfit.pro \citep{Mar2009}, PHOENIX (\citealt{All2012}; \citealt{Bar2015}), amoeba.pro, jktld \citep{Sou2015}}

\clearpage

%%%Here are all of the Figures

 \clearpage

\clearpage

\clearpage

\clearpage

 \clearpage

\clearpage

\clearpage

\clearpage

\clearpage

%%$$$$$$$$ TABLES

\begin{deluxetable}{llrrrr}
\tablecaption{Description of APOGEE M Dwarf Sample\label{tab:sample}}
\tablenum{1}

\tablehead{\colhead{\teff} & \colhead{N} & \colhead{H} & \colhead{J} & \colhead{K} & \colhead{SNR\tablenotemark{a}} \\ 
\colhead{(K)} & \colhead{} & \colhead{(mag)} & \colhead{(mag)} & \colhead{(mag)} & \colhead{} }
\startdata
$2600 - 2700$ & $5$ & $10.4 - 11.8$ & $11.1 - 12.4$ & $10.0 - 11.4$ & $108 - 217$ \\
$2700 - 2800$ & $9$ & $7.19 - 12.5$ & $7.79 - 13.1$ & $6.85 - 12.1$ & $65.0 - 567$ \\
$2800 - 2900$ & $7$ & $10.5 - 12.4$ & $11.1 - 13.0$ & $10.2 - 12.1$ & $58.4 - 264$ \\
$2900 - 3000$ & $12$ & $9.49 - 12.5$ & $10.0 - 13.1$ & $9.17 - 12.2$ & $50.9 - 589$ \\
$3000 - 3100$ & $35$ & $9.16 - 13.2$ & $9.75 - 13.8$ & $8.89 - 13.0$ & $51.1 - 493$ \\
$3100 - 3200$ & $58$ & $8.35 - 12.3$ & $8.87 - 12.9$ & $8.05 - 12.0$ & $51.5 - 773$ \\
$3200 - 3300$ & $106$ & $8.05 - 12.3$ & $8.59 - 12.9$ & $7.77 - 12.0$ & $50.5 - 1470$ \\
$3300 - 3400$ & $139$ & $7.03 - 12.4$ & $7.58 - 12.9$ & $6.81 - 12.2$ & $64.1 - 985$ \\
$3400 - 3500$ & $105$ & $7.56 - 12.0$ & $8.12 - 12.6$ & $7.32 - 11.8$ & $50.3 - 1390$ \\
$3500 - 3600$ & $72$ & $7.12 - 11.9$ & $7.70 - 12.5$ & $6.89 - 11.7$ & $54.8 - 1740$ \\
$3600 - 3700$ & $70$ & $7.03 - 12.0$ & $7.57 - 12.5$ & $6.77 - 11.7$ & $50.9 - 1100$ \\
$3700 - 3800$ & $43$ & $7.32 - 11.8$ & $7.89 - 12.4$ & $7.09 - 11.6$ & $58.0 - 1150$ \\
$3800 - 3900$ & $41$ & $7.26 - 11.6$ & $7.85 - 12.3$ & $7.04 - 11.5$ & $57.5 - 1170$ \\
$3900 - 4000$ & $14$ & $7.37 - 11.8$ & $7.96 - 12.5$ & $7.18 - 11.7$ & $146 - 1190$ \\
\hline \\
$2609 - 3999$ & $714$ & $7.03 - 13.2$ & $7.57 - 13.8$ & $6.77 - 13.0$ & $50.3 - 1740$ \\
\enddata

\tablenotetext{a}{According to the APOGEE documentation, due to unquantified systematic errors, the maximum SNR is likely limited to $\sim200$.  The SNR quoted here is based on the statistical flux error estimates, which do not necessarily reflect the SNR limit.}

\end{deluxetable}

\clearpage

\begin{deluxetable}{lcc} 
\tablecaption{PHOENIX template grid parameters\label{tab:grid_pars}}
\tablenum{2}

\tablehead{\colhead{Quantity} & \colhead{Range} & \colhead{Increment} \\ 
\colhead{} & \colhead{} & \colhead{} } 

\startdata
\logg & $4.5 \textrm{--} \unit[5.5]{dex}$ & $\unit[0.5]{dex}$ \\
\teff & $2600 \textrm{--} \unit[4000]{K}$ & $\unit[100]{K}$ \\
\met & $\unit[-2.5]{}~ \textrm{to} \unit[+0.5]{dex}$ & varies \\
\vsini & $1 \textrm{--} \unit[100]{\kms}$ & varies \\
\enddata

\end{deluxetable}

\clearpage

\begin{deluxetable}{llccrlrrr}

\tablecaption{Comparison to \vsinie Measurements from the Literature\label{table:vsini_overlap_tab}}
\tabletypesize{\footnotesize}
\tablenum{3}
\tablehead{
\colhead{2MASS ID} &
\colhead{Name} &
\colhead{RA} &
\colhead{DEC} &
\colhead{\teff} &
\colhead{Spectral} & 
\colhead{$\vsini_{\textrm{VFIT}}$} &
\colhead{$\vsini_{\textrm{LIT}}$} &
\colhead{Resolution} \\
\colhead{} &
\colhead{} &
\colhead{(deg. J2000)} &
\colhead{(deg. J2000)} &
\colhead{(K)} &
\colhead{Type\tablenotemark{a}} &
\colhead{(\kms)} &
\colhead{(\kms)} &
\colhead{(R/1000)} }
\startdata
J02085359$+$4926565 & GJ 3136 & $32.223315$ & $49.449055$ & $3340$ & M4.0V & $22.9\pm1.5$ & $30.0 \pm{5.0}^{1} $  & $19$ \\
J03212176$+$7958022 & GJ 133 & $50.340691$ & $79.967285$ & $3586$ & M2.0Ve & $< 8.0$ & $< 1.0^{2}$ & $75/115$ \\
J04584599$+$5056378 & GJ 1074 & $74.691634$ & $50.943859$ & $3807$ & M1.0Ve & $< 8.0$ & $< 4.0^{3}$ & $40/48$ \\
J05470907$-$0512106 & LHS 1785 & $86.787800$ & $-5.2029690$ & $3149$ & M4.5V & $< 8.0$ & $4.50 \pm{0.60}^{4} $ & $37$ \\
J06421118$+$0334527 & G 108-21 & $100.54659$ & $3.5813060$ & $3437$ & M3V & $< 8.0$ & $0.900^{2} $ & $75/115$ \\
J09005033$+$0514293 & Ross 687 & $135.20971$ & $5.2414950$ & $3392$ & M3.0Ve & $< 8.0$ & $< 3.0^{5}$ & $40$ \\
J09422327$+$5559015 & GJ 363 & $145.59698$ & $55.983776$ & $3390$ & M3V & $< 8.0$ & $< 3.0^{5}$ & $40$ \\
J10355725$+$2853316 & UCAC4 595-047332 & $158.98856$ & $28.892134$ & $3442$ & M3.0V & $< 8.0$ & $4.00 \pm{2.0}^{6} $ & $50$ \\
J13085124$-$0131075 & \nodata & $197.21351$ & $-1.5187690$ & $3498$ & M3.0V & $< 8.0$ & $< 2.0^{9}$ & $65/68$ \\
J13455527$+$2723131 & LHS 2795 & $206.48032$ & $27.386990$ & $3318$ & \nodata  & $< 8.0$ & $6.40 \pm{1.0}^{2} $& $75/115$ \\
J13564148$+$4342587 & LP 220-13 & $209.17285$ & $43.716324$ & $2626$ & M8V & $13.5\pm1.6$ & $14.0 \pm{2.0}^{7} $ & $31/32$ \\
J14333985$+$0920094 & HD 127871B & $218.41606$ & $9.3359630$ & $3384$ & M3.5V & $< 8.0$ & $5.30 \pm{1.0}^{2} $ & $75/115$ \\
J14573227$+$3123446 & Ross 53 & $224.38448$ & $31.395733$ & $3884$ & K5V & $< 8.0$ & $2.63 \pm{1.0}^{2} $ & $75/115$ \\
J16404891$+$3618596 & Ross 812 & $250.20383$ & $36.316566$ & $3662$ & M2V & $< 8.0$ & $< 4.0^{3}$ & $40/48$ \\
J19454969$+$3223132 & LP 337-3 & $296.45707$ & $32.387005$ & $3623$ & M1.5Ve & $< 8.0$ & $< 4.0^{3}$ & $40/48$ \\
J19510930$+$4628598 & GJ 1243 & $297.78877$ & $46.483295$ & $3205$ & M4.0V & $22.5\pm1.6$ & $22.0 \pm{3.0}^{3} $ & $40/48$ \\
J19535443$+$4424541 & G 208-44 & $298.47680$ & $44.415043$ & $2749$ & M5.5Ve & $22.6\pm1.5$ & $22.5^{8} $ & $31$ \\
\enddata
\tablerefs{(1) \cite{Giz2002}\,[G02]; (2) \cite{Hou2015}\,[HM15]; (3) \cite{Rei2012}\,[R12]; (4) \cite{Jen2009}\,[J09]; (5) \cite{DavT2015}\,[D15]; (6) \cite{Sch2012}\,[S12]; (7) \cite{RB2010}\,[RB10]; (8) \cite{MB2003}\,[MB03]; (9) \cite{Mou2017}\,[M17]}
\tablenotetext{a}{From SIMBAD database \citep{Wen2000}}

\end{deluxetable}

\clearpage

\begin{deluxetable}{lccccccr}
\tablecaption{\vsinie Results\label{table:myvsini}}
\tablenum{4}
\tablehead{\colhead{2MASS ID} &
\colhead{RA} &
\colhead{DEC} &
\colhead{\taspcap} &
\colhead{T$_{\textrm{eff, VFIT}}$} &
\colhead{$\met_{\textrm{VFIT}}$} &
\colhead{$\logg_{\textrm{VFIT}}$} &
\colhead{$\vsini_{\textrm{VFIT}}$} \\
\colhead{} & 
\colhead{(degrees J2000)} &
\colhead{(degrees J2000)} &
\colhead{(K)} &
\colhead{(K)} &
\colhead{(dex)} &
\colhead{(dex)} &
\colhead{(\kms)}} 
\startdata
J00004701$+$1624101 & $0.19587600$ & $16.402811$ & $3725$ & $3800$ & $+0.5$ & $5.0$ & $<8.0$\\
J00034394$+$8606422 & $0.93308800$ & $86.111732$ & $2893$ & $2800$ & $-0.0$ & $5.0$ & $13.2\pm1.5$\\
J00255540$+$5749320 & $6.4808560$ & $57.825562$ & $3400$ & $3300$ & $+0.5$ & $5.0$ & $<8.0$\\
J00255888$+$5559296 & $6.4953360$ & $55.991581$ & $3589$ & $3500$ & $-0.0$ & $4.5$ & $11.5\pm1.5$\\
J00262872$+$6747026 & $6.6196670$ & $67.784073$ & $3756$ & $3800$ & $-0.0$ & $4.5$ & $<8.0$\\
J00270673$+$4941531 & $6.7780790$ & $49.698093$ & $3297$ & $3300$ & $-0.0$ & $5.0$ & $<8.0$\\
J00285391$+$5022330 & $7.2246660$ & $50.375839$ & $3236$ & $3200$ & $+0.5$ & $5.0$ & $14.2\pm1.5$\\
J00301250$+$5028392 & $7.5520980$ & $50.477570$ & $3209$ & $3200$ & $-0.0$ & $5.0$ & $<8.0$\\
J02081218$+$1508424 & $32.050754$ & $15.145118$ & $3174$ & $3200$ & $+0.5$ & $5.5$ & $<8.0$\\
J02081366$+$4949023 & $32.056958$ & $49.817318$ & $2824$ & $2600$ & $-0.0$ & $5.0$ & $<8.0$\\
J02085359$+$4926565 & $32.223315$ & $49.449055$ & $3340$ & $3300$ & $+0.5$ & $5.0$ & $22.9\pm1.5$\\
J02122001$+$1249287 & $33.083411$ & $12.824662$ & $3459$ & $3700$ & $+0.5$ & $5.5$ & $<8.0$\\
J02144781$+$5334438 & $33.699241$ & $53.578857$ & $3333$ & $3400$ & $-0.5$ & $5.0$ & $<8.0$\\
J05320969$+$2754534 & $83.040394$ & $27.914845$ & $2865$ & $2900$ & $-0.0$ & $5.5$ & $8.36\pm1.7$\\
J05325989$+$2608271 & $83.249549$ & $26.140879$ & $3342$ & $3400$ & $+0.5$ & $5.5$ & $47.7\pm2.3$\\
\enddata
\tablecomments{This table is available in its entirety in a machine-readable form in the online journal. A portion is shown here for guidance regarding its form and content.}

\end{deluxetable}

\clearpage

\end{document}